\documentclass[numbers,twocolumn]{bmcart}
\makeatletter
\def\ps@pprintTitle{%
 \let\@oddhead\@empty
 \let\@evenhead\@empty
 \def\@oddfoot{}%
 \let\@evenfoot\@oddfoot}
\makeatother

\RequirePackage[authoryear]{natbib}
\usepackage{amssymb}
\usepackage{amsmath}
\usepackage{lineno}
\usepackage{hyperref} 
\usepackage{enumitem}
\usepackage{titlesec}
\usepackage{ragged2e}
\usepackage{graphicx}

\hypersetup{
    colorlinks,
}

\usepackage{fancyvrb}
\usepackage{color}
\usepackage{verbatim}

\makeatletter

\startlocaldefs

\def\PY@reset{\let\PY@it=\relax \let\PY@bf=\relax%
    \let\PY@ul=\relax \let\PY@tc=\relax%
    \let\PY@bc=\relax \let\PY@ff=\relax}
\def\PY@tok#1{\csname PY@tok@#1\endcsname}
\def\PY@toks#1+{\ifx\relax#1\empty\else%
    \PY@tok{#1}\expandafter\PY@toks\fi}
\def\PY@do#1{\PY@bc{\PY@tc{\PY@ul{%
    \PY@it{\PY@bf{\PY@ff{#1}}}}}}}
\def\PY#1#2{\PY@reset\PY@toks#1+\relax+\PY@do{#2}}

\expandafter\def\csname PY@tok@gd\endcsname{\def\PY@tc##1{\textcolor[rgb]{0.63,0.00,0.00}{##1}}}
\expandafter\def\csname PY@tok@gu\endcsname{\let\PY@bf=\textbf\def\PY@tc##1{\textcolor[rgb]{0.50,0.00,0.50}{##1}}}
\expandafter\def\csname PY@tok@gt\endcsname{\def\PY@tc##1{\textcolor[rgb]{0.00,0.27,0.87}{##1}}}
\expandafter\def\csname PY@tok@gs\endcsname{\let\PY@bf=\textbf}
\expandafter\def\csname PY@tok@gr\endcsname{\def\PY@tc##1{\textcolor[rgb]{1.00,0.00,0.00}{##1}}}
\expandafter\def\csname PY@tok@cm\endcsname{\let\PY@it=\textit\def\PY@tc##1{\textcolor[rgb]{0.25,0.50,0.50}{##1}}}
\expandafter\def\csname PY@tok@vg\endcsname{\def\PY@tc##1{\textcolor[rgb]{0.10,0.09,0.49}{##1}}}
\expandafter\def\csname PY@tok@m\endcsname{\def\PY@tc##1{\textcolor[rgb]{0.40,0.40,0.40}{##1}}}
\expandafter\def\csname PY@tok@mh\endcsname{\def\PY@tc##1{\textcolor[rgb]{0.40,0.40,0.40}{##1}}}
\expandafter\def\csname PY@tok@go\endcsname{\def\PY@tc##1{\textcolor[rgb]{0.53,0.53,0.53}{##1}}}
\expandafter\def\csname PY@tok@ge\endcsname{\let\PY@it=\textit}
\expandafter\def\csname PY@tok@vc\endcsname{\def\PY@tc##1{\textcolor[rgb]{0.10,0.09,0.49}{##1}}}
\expandafter\def\csname PY@tok@il\endcsname{\def\PY@tc##1{\textcolor[rgb]{0.40,0.40,0.40}{##1}}}
\expandafter\def\csname PY@tok@cs\endcsname{\let\PY@it=\textit\def\PY@tc##1{\textcolor[rgb]{0.25,0.50,0.50}{##1}}}
\expandafter\def\csname PY@tok@cp\endcsname{\def\PY@tc##1{\textcolor[rgb]{0.74,0.48,0.00}{##1}}}
\expandafter\def\csname PY@tok@gi\endcsname{\def\PY@tc##1{\textcolor[rgb]{0.00,0.63,0.00}{##1}}}
\expandafter\def\csname PY@tok@gh\endcsname{\let\PY@bf=\textbf\def\PY@tc##1{\textcolor[rgb]{0.00,0.00,0.50}{##1}}}
\expandafter\def\csname PY@tok@ni\endcsname{\let\PY@bf=\textbf\def\PY@tc##1{\textcolor[rgb]{0.60,0.60,0.60}{##1}}}
\expandafter\def\csname PY@tok@nl\endcsname{\def\PY@tc##1{\textcolor[rgb]{0.63,0.63,0.00}{##1}}}
\expandafter\def\csname PY@tok@nn\endcsname{\let\PY@bf=\textbf\def\PY@tc##1{\textcolor[rgb]{0.00,0.00,1.00}{##1}}}
\expandafter\def\csname PY@tok@no\endcsname{\def\PY@tc##1{\textcolor[rgb]{0.53,0.00,0.00}{##1}}}
\expandafter\def\csname PY@tok@na\endcsname{\def\PY@tc##1{\textcolor[rgb]{0.49,0.56,0.16}{##1}}}
\expandafter\def\csname PY@tok@nb\endcsname{\def\PY@tc##1{\textcolor[rgb]{0.00,0.50,0.00}{##1}}}
\expandafter\def\csname PY@tok@nc\endcsname{\let\PY@bf=\textbf\def\PY@tc##1{\textcolor[rgb]{0.00,0.00,1.00}{##1}}}
\expandafter\def\csname PY@tok@nd\endcsname{\def\PY@tc##1{\textcolor[rgb]{0.67,0.13,1.00}{##1}}}
\expandafter\def\csname PY@tok@ne\endcsname{\let\PY@bf=\textbf\def\PY@tc##1{\textcolor[rgb]{0.82,0.25,0.23}{##1}}}
\expandafter\def\csname PY@tok@nf\endcsname{\def\PY@tc##1{\textcolor[rgb]{0.00,0.00,1.00}{##1}}}
\expandafter\def\csname PY@tok@si\endcsname{\let\PY@bf=\textbf\def\PY@tc##1{\textcolor[rgb]{0.73,0.40,0.53}{##1}}}
\expandafter\def\csname PY@tok@s2\endcsname{\def\PY@tc##1{\textcolor[rgb]{0.73,0.13,0.13}{##1}}}
\expandafter\def\csname PY@tok@vi\endcsname{\def\PY@tc##1{\textcolor[rgb]{0.10,0.09,0.49}{##1}}}
\expandafter\def\csname PY@tok@nt\endcsname{\let\PY@bf=\textbf\def\PY@tc##1{\textcolor[rgb]{0.00,0.50,0.00}{##1}}}
\expandafter\def\csname PY@tok@nv\endcsname{\def\PY@tc##1{\textcolor[rgb]{0.10,0.09,0.49}{##1}}}
\expandafter\def\csname PY@tok@s1\endcsname{\def\PY@tc##1{\textcolor[rgb]{0.73,0.13,0.13}{##1}}}
\expandafter\def\csname PY@tok@kd\endcsname{\let\PY@bf=\textbf\def\PY@tc##1{\textcolor[rgb]{0.00,0.50,0.00}{##1}}}
\expandafter\def\csname PY@tok@sh\endcsname{\def\PY@tc##1{\textcolor[rgb]{0.73,0.13,0.13}{##1}}}
\expandafter\def\csname PY@tok@sc\endcsname{\def\PY@tc##1{\textcolor[rgb]{0.73,0.13,0.13}{##1}}}
\expandafter\def\csname PY@tok@sx\endcsname{\def\PY@tc##1{\textcolor[rgb]{0.00,0.50,0.00}{##1}}}
\expandafter\def\csname PY@tok@bp\endcsname{\def\PY@tc##1{\textcolor[rgb]{0.00,0.50,0.00}{##1}}}
\expandafter\def\csname PY@tok@c1\endcsname{\let\PY@it=\textit\def\PY@tc##1{\textcolor[rgb]{0.25,0.50,0.50}{##1}}}
\expandafter\def\csname PY@tok@kc\endcsname{\let\PY@bf=\textbf\def\PY@tc##1{\textcolor[rgb]{0.00,0.50,0.00}{##1}}}
\expandafter\def\csname PY@tok@c\endcsname{\let\PY@it=\textit\def\PY@tc##1{\textcolor[rgb]{0.25,0.50,0.50}{##1}}}
\expandafter\def\csname PY@tok@mf\endcsname{\def\PY@tc##1{\textcolor[rgb]{0.40,0.40,0.40}{##1}}}
\expandafter\def\csname PY@tok@err\endcsname{\def\PY@bc##1{\setlength{\fboxsep}{0pt}\fcolorbox[rgb]{1.00,0.00,0.00}{1,1,1}{\strut ##1}}}
\expandafter\def\csname PY@tok@mb\endcsname{\def\PY@tc##1{\textcolor[rgb]{0.40,0.40,0.40}{##1}}}
\expandafter\def\csname PY@tok@ss\endcsname{\def\PY@tc##1{\textcolor[rgb]{0.10,0.09,0.49}{##1}}}
\expandafter\def\csname PY@tok@sr\endcsname{\def\PY@tc##1{\textcolor[rgb]{0.73,0.40,0.53}{##1}}}
\expandafter\def\csname PY@tok@mo\endcsname{\def\PY@tc##1{\textcolor[rgb]{0.40,0.40,0.40}{##1}}}
\expandafter\def\csname PY@tok@kn\endcsname{\let\PY@bf=\textbf\def\PY@tc##1{\textcolor[rgb]{0.00,0.50,0.00}{##1}}}
\expandafter\def\csname PY@tok@mi\endcsname{\def\PY@tc##1{\textcolor[rgb]{0.40,0.40,0.40}{##1}}}
\expandafter\def\csname PY@tok@gp\endcsname{\let\PY@bf=\textbf\def\PY@tc##1{\textcolor[rgb]{0.00,0.00,0.50}{##1}}}
\expandafter\def\csname PY@tok@o\endcsname{\def\PY@tc##1{\textcolor[rgb]{0.40,0.40,0.40}{##1}}}
\expandafter\def\csname PY@tok@kr\endcsname{\let\PY@bf=\textbf\def\PY@tc##1{\textcolor[rgb]{0.00,0.50,0.00}{##1}}}
\expandafter\def\csname PY@tok@s\endcsname{\def\PY@tc##1{\textcolor[rgb]{0.73,0.13,0.13}{##1}}}
\expandafter\def\csname PY@tok@kp\endcsname{\def\PY@tc##1{\textcolor[rgb]{0.00,0.50,0.00}{##1}}}
\expandafter\def\csname PY@tok@w\endcsname{\def\PY@tc##1{\textcolor[rgb]{0.73,0.73,0.73}{##1}}}
\expandafter\def\csname PY@tok@kt\endcsname{\def\PY@tc##1{\textcolor[rgb]{0.69,0.00,0.25}{##1}}}
\expandafter\def\csname PY@tok@ow\endcsname{\let\PY@bf=\textbf\def\PY@tc##1{\textcolor[rgb]{0.67,0.13,1.00}{##1}}}
\expandafter\def\csname PY@tok@sb\endcsname{\def\PY@tc##1{\textcolor[rgb]{0.73,0.13,0.13}{##1}}}
\expandafter\def\csname PY@tok@k\endcsname{\let\PY@bf=\textbf\def\PY@tc##1{\textcolor[rgb]{0.00,0.50,0.00}{##1}}}
\expandafter\def\csname PY@tok@se\endcsname{\let\PY@bf=\textbf\def\PY@tc##1{\textcolor[rgb]{0.73,0.40,0.13}{##1}}}
\expandafter\def\csname PY@tok@sd\endcsname{\let\PY@it=\textit\def\PY@tc##1{\textcolor[rgb]{0.73,0.13,0.13}{##1}}}

\def\PYZus{\char`\_}
\def\PYZob{\char`\{}
\def\PYZcb{\char`\}}

\def\PYZgt{\char`\>}
\def\PYZsh{\char`\#}

\def\PYZhy{\char`\-}
\def\PYZsq{\char`\'}
\def\PYZdq{\char`\"}

\makeatother

\hypersetup{
  pdfinfo={
    Title={The Illustris Simulation: Public Data Release},
    Author={Dylan Nelson},
    Subject={},
    Keywords={}
  }
  pdfcreator={},
  pdfproducer={}
}

\endlocaldefs

\begin{document}

\begin{frontmatter}

\begin{fmbox}
\dochead{Research}


\title{The IllustrisTNG Simulations: \\Public Data Release\texorpdfstring{$^\dagger$}{'}}

\author[addressref={mpa},email={dnelson@mpa-garching.mpg.de}]{Dylan Nelson}
\author[addressref={mpa,hits,heidelberg}]{Volker Springel}
\author[addressref={mpia}]{Annalisa Pillepich}
\author[addressref={unam}]{Vicente Rodriguez-Gomez}
\author[addressref={florida,mit}]{Paul Torrey}
\author[addressref={cca}]{Shy Genel}
\author[addressref={mit}]{Mark Vogelsberger}
\author[addressref={mpa,hits}]{Ruediger Pakmor}
\author[addressref={bologna,mit}]{Federico Marinacci}
\author[addressref={cfa}]{Rainer Weinberger}
\author[addressref={northwestern}]{Luke Kelley}
\author[addressref={iceland,durham}]{Mark Lovell}
\author[addressref={cfa}]{Benedikt Diemer}
\author[addressref={cfa}]{Lars Hernquist}

\vspace{0.5em}$\dagger$ Permanently available at \href{http://www.tng-project.org/data/}{www.tng-project.org/data}

\address[id=mpa]{Max-Planck-Institut f\"{u}r Astrophysik, Karl-Schwarzschild Str. 1, 85741 Garching, Germany}
\address[id=mpia]{Max-Planck-Institut f\"{u}r Astronomie, K\"{o}nigstuhl 17, 69117 Heidelberg, Germany}
\address[id=cfa]{Center for Astrophysics, Harvard \& Smithsonian, 60 Garden Street, Cambridge, MA, 02138, USA}
\address[id=cca]{Center for Computational Astrophysics, Flatiron Institute, 162 Fifth Avenue, New York, NY 10010, USA}
\address[id=mit]{Kavli Institute for Astrophysics and Space Research, Department of Physics, MIT, Cambridge, MA, 02139, USA}
\address[id=hits]{Heidelberg Institute for Theoretical Studies, Schloss-Wolfsbrunnenweg 35, 69118 Heidelberg, Germany}
\address[id=heidelberg]{Zentrum f\"{u}r Astronomie der Universit\"{a}t Heidelberg, ARI, M\"{o}nchhofstr. 12-14, 69120 Heidelberg, Germany}
\address[id=unam]{Instituto de Radioastronom\'{i}a y Astrof\'{i}sica, Universidad Nacional Aut\'{o}noma de M\'{e}xico, Apdo. Postal 72-3, 58089 Morelia, Mexico}
\address[id=florida]{University of Florida, Department of Physics, 2001 Museum Rd., Gainesville, FL 32611, USA}
\address[id=bologna]{Department of Physics and Astronomy, University of Bologna, Piero Gobetti 93/2, I-40129 Bologna, Italy}
\address[id=northwestern]{Center for Interdisciplinary Exploration and Research in Astrophysics, Northwestern University, Evanston, IL 60208, USA}
\address[id=iceland]{Center for Astrophysics and Cosmology, Science Institute, University of Iceland, Dunhagi 5, 107 Reykjavik, Iceland}
\address[id=durham]{Institute for Computational Cosmology, Durham University, South Road, Durham DH1 3LE, UK}

\end{fmbox}

\begin{abstractbox}

\begin{abstract}
\justifying
We present the full public release of all data from the TNG50, TNG100 and TNG300 simulations of the IllustrisTNG project. IllustrisTNG is a suite of large volume, cosmological, gravo-magnetohydrodynamical simulations run with the moving-mesh code {\sc Arepo}. TNG includes a comprehensive model for galaxy formation physics, and each TNG simulation self-consistently solves for the coupled evolution of dark matter, cosmic gas, luminous stars, and supermassive blackholes from early time to the present day, $z=0$. Each of the flagship runs -- TNG50, TNG100, and TNG300 -- are accompanied by halo/subhalo catalogs, merger trees, lower-resolution and dark-matter only counterparts, all available with 100 snapshots. We discuss scientific and numerical cautions and caveats relevant when using TNG. 

The data volume now directly accessible online is $\sim$1.1 PB, including 2,000 full volume snapshots and $\sim$115,000 high time-resolution subbox snapshots. Data access and analysis examples are available in IDL, Python, and Matlab. We describe improvements and new functionality in the web-based API, including on-demand visualization and analysis of galaxies and halos, exploratory plotting of scaling relations and other relationships between galactic and halo properties, and a new JupyterLab interface. This provides an online, browser-based, near-native data analysis platform enabling user computation with local access to TNG data, alleviating the need to download large datasets. 
\end{abstract}


\end{abstractbox}

\end{frontmatter}


\section*{Main Text}

\section{Introduction}

Some of our most powerful tools for understanding the origin and evolution of large-scale cosmic structure and the galaxies which form therein are cosmological simulations. From pioneering beginnings \citep{ps74,davis85}, dark matter, gravity-only simulations have evolved into cosmological hydrodynamical simulations \citep{katz92}. These aim to self-consistently model the coupled evolution of dark matter, gas, stars, and blackholes at a minimum, and are now being extended to also include magnetic fields, radiation, cosmic rays, and other fundamental physical components. Such simulations provide foundational support in our understanding of the $\Lambda$CDM cosmological model, including the nature of both dark matter and dark energy. 

Modern large-volume simulations now capture cosmological scales of tens to hundreds of comoving megaparsecs, while simultaneously resolving the internal structure of individual galaxies at $\lesssim\,$1 kpc scales. Recent examples reaching $z=0$ include Illustris \citep{vog14a,genel14}, EAGLE \citep{schaye15,crain15}, Horizon-AGN \citep{dubois14}, Romulus \citep{tremmel17}, Simba \citep{dave19}, Magneticum \citep{dolag16}, among others. These simulations produce observationally verifiable outcomes across a diverse range of astrophysical regimes, from the stellar and gaseous properties of galaxies, galaxy populations, and the supermassive blackholes they host, to the expected distribution of molecular, neutral, and ionized gas tracers across interstellar, circumgalactic, and intergalactic scales, in addition to the expected distribution of the dark matter component itself. 

Complementary efforts, although not the focus of this data release, include high redshift reionization-era simulations such as BlueTides \citep{feng16}, Sphinx \citep{rosdahl18}, and CoDa II \citep{ocvirk19}, among others. In addition, `zoom' simulation campaigns include NIHAO \citep{wang15}, FIRE-2 \citep{hopkins18}, and Auriga \citep{grand17}, in addition to many others. These have provided numerous additional insights into many questions in galaxy evolution \citep[recent progress reviewed in][]{fg18}. For instance, reionization simulations may be able to include explicit radiative transfer, and zoom simulations may be able to reach higher resolutions and/or more rapidly explore model variations, in comparison to large cosmological volume simulations.

Observational efforts studying the properties of galaxies across cosmic time provide ever richer datasets. Surveys such as SDSS \citep{york00}, CANDELS \citep{grogin11}, 3D-HST \citep{brammer12}, LEGA-C \citep{vanderwel16}, SINS/zC-SINF and KMOS3D \citep{genzel14,wisnioski15}, KBSS \citep{steidel14}, and MOSDEF \citep{kriek15} provide local and high redshift measurements of the statistical properties of galaxy populations. Complementary, spatially-resolved data has recently become available from large, $z=0$ IFU surveys such as MANGA \citep{bundy15}, CALIFA \citep{sanchez12} and SAMI \citep{bryant15}.

In order to inform theoretical models using observational constraints, as well as to interpret observational results using realistic cosmological models, public data dissemination from both observational and simulation campaigns is required. Observational data release has a successful history dating back at least to the SDSS SkyServer \citep{szalay00,szalay02a}, which provides tools for remote users to query and acquire large datasets \citep{gray02,szalay02b}. The still-in-use approach is based on user written SQL queries, which provide search results as well as data acquisition. From the theoretical community, the public data release of the Millennium simulation \citep{spr05c} was the first attempt of similar scale. Modeled on the SDSS approach, data was stored in a relational database, with an immediately recognizable SQL-query interface \citep{lemson06}. It has since been extended to include additional simulations, including Millennium-II \citep{boylan09,guo11}, and a first attempt at the idea of a ``virtual observatory'' (VO) was realized \citep{overzier13}. The Theoretical Astrophysical Observatory \citep[TAO;][]{bernyk14} was similarly focused around mock observations of simulated galaxy and galaxy survey data. Explorations continue on how to best deliver theoretical results within the existing VO framework \citep{lemson09,lemson14}. 

Other dark-matter only simulations have released data with similar approaches, including Bolshoi and MultiDark \citep[CosmoSim;][]{klypin11,riebe13}, DEUS \citep{rasera10}, and MICE \citep[Cosmohub;][]{crocce10}. In contrast, some recent simulation projects have made group catalogs and/or snapshots available for direct download, including MassiveBlack-II \citep{khandai14}, the Dark Sky simulation \citep{skillman14}, $\nu^2$GC \citep{makiya16}, and Abacus \citep{lehman18}. Skies and Universes \citep{klypin18} organizes a number of such data releases. With respect to Illustris, the most comparable in simulation type, data complexity, and scientific scope is the recent public data release of the Eagle simulation, described in \cite{mcalpine15} \citep[see also][]{camps18}. The initial group catalog release was modeled on the Millennium database architecture, and the raw snapshot data was also subsequently made available \citep{eagle17}. More recently, significant infrastructure research and development has focused on providing remote computational resources, including the NOAO Data Lab \citep{fitzpatrick14} and the SciServer project \citep{medvedev16,raddick17}. Web-based orchestration projects also include \cite{ragagnin17}, Tangos \citep{pontzen18}, and Jovial \citep{araya18}.

\begin{figure*}[tb!]
\centerline{\includegraphics[angle=0,width=7.0in]{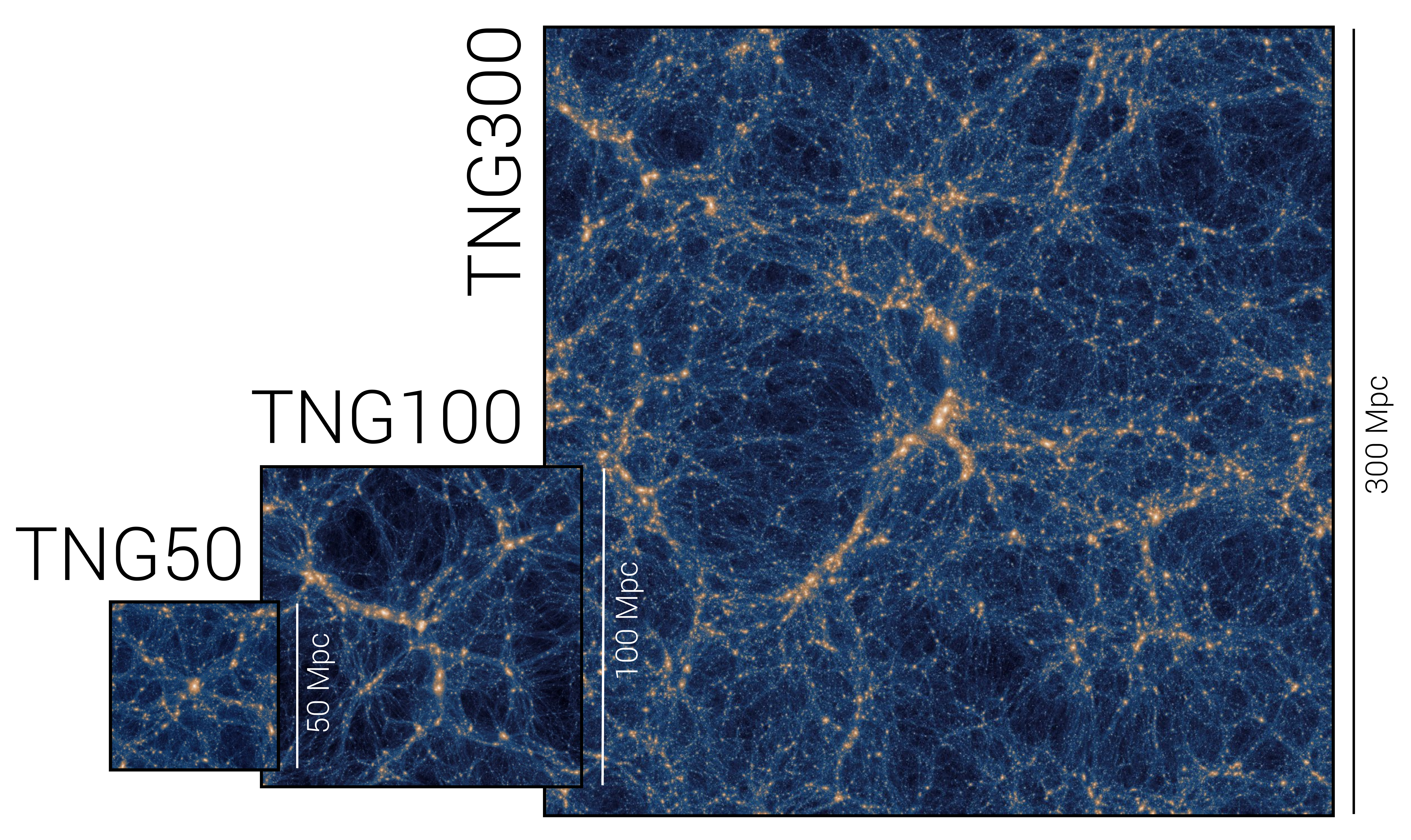}}
\caption{ The three IllustrisTNG simulation volumes: TNG50, TNG100, and TNG300, shown here in projected dark matter density. In each case the name denotes the box side-length in comoving Mpc. The largest, TNG300, enables the study of rare, massive objects such as galaxy clusters, and provides unparalleled statistics of the galaxy population as a whole. TNG50, with a mass resolution more than one hundred times better, provides for the detailed examination of internal, structural properties and small-scale phenomena. In the middle, TNG100 uses the same initial conditions as the original Illustris simulation, providing a useful balance of resolution and volume for studying many aspects of galaxy evolution.
 \label{fig_3boxes}} 
\end{figure*}

The public release of IllustrisTNG (hereafter, TNG) follows upon and further develops tools and ideas pioneered in the original Illustris data release. We offer direct online access to all snapshot, group catalog, merger tree, and supplementary data catalog files. In addition, we develop a web-based API which allows users to perform many common tasks without the need to download any full data files. These include searching over the group catalogs, extracting particle data from the snapshots, accessing individual merger trees, and requesting visualization and further data analysis functions. Extensive documentation and programmatic examples (in the IDL, Python, and Matlab languages) are provided.

This paper is intended primarily as an overview guide for TNG data users, describing updates and new features, while exhaustive documentation will be maintained online. In Section \ref{sSims} we give an overview of the simulations. Section \ref{sDataProducts} describes the data products, and Section \ref{sDataAccess} discusses methods for data access. In Section \ref{sRemarks} we present some scientific remarks and cautions, while in Section \ref{sCommunity} we discuss community considerations including citation requests. Section \ref{sImplementation} describes technical details related to the data release architecture, while Section \ref{sConclusions} summarizes. Appendix A provides a few additional data details, while Appendix B shows several examples of how to use the API.


\section{Description of the Simulations} \label{sSims}

IllustrisTNG is a suite of large volume, cosmological, gravo-magnetohydrodynamical simulations run with the moving-mesh code {\textsc AREPO} \citep{spr10}. The TNG project is made up of three simulation volumes: TNG50, TNG100, and TNG300. The first two simulations, TNG100 and TNG300, were recently introduced in a series of five presentation papers \citep{springel18,pillepich18b,naiman18,nelson18a,marinacci18}, and these are here publicly released in full. The third and final simulation of the project is TNG50 \citep{pillepich19,nelson19b}, with substantially higher resolution. TNG includes a comprehensive model for galaxy formation physics, which is able to realistically follow the formation and evolution of galaxies across cosmic time \citep{weinberger17,pillepich18a}. Each TNG simulation solves for the coupled evolution of dark matter, cosmic gas, luminous stars, and supermassive blackholes from a starting redshift of $z=127$ to the present day, $z=0$.

The IllustrisTNG project\footnote{\url{www.tng-project.org}} is the successor of the original Illustris simulation\footnote{\url{www.illustris-project.org}} \citep{vog14a,vog14b,genel14,sijacki15} and its associated galaxy formation model \citep{vog13,torrey14}. Illustris was publicly released in its entirety roughly three and a half years ago \citep{nelson15b}. TNG incorporates an updated `next generation' galaxy formation model which includes new physics and numerical improvements, as well as refinements to the original model. TNG newly includes a treatment of cosmic magnetism, following the amplification and dynamical impact of magnetic fields, as described below.

The three flagship runs of IllustrisTNG are each accompanied by lower-resolution and dark-matter only counterparts. Three physical simulation box sizes are employed: cubic volumes of roughly 50, 100, and 300 Mpc side length, which we refer to as TNG50, TNG100, and TNG300, respectively. The three boxes complement each other by enabling investigations of various aspects of galaxy formation. The large physical volume associated with the largest simulation box (TNG300) enables, for instance, the study of galaxy clustering, the analysis of rare and massive objects such as galaxy clusters, and provides the largest statistical galaxy sample. In contrast, the smaller physical volume simulation of TNG50 enables a mass resolution which is more than a hundred times better than in the TNG300 simulation, providing a more detailed look at, for example, the structural properties of galaxies, and small-scale gas phenomena in and around galaxies. Situated in the middle, the TNG100 simulation uses the same initial conditions (identical phases, adjusted for the updated cosmology) as the original Illustris simulation. This facilitates robust comparisons between the original Illustris results and the updated TNG model. For many galaxy evolution analyses, TNG100 provides an ideal balance of volume and resolution, particularly for intermediate mass halos. Despite these strengths, each volume still has intrinsic physical and numerical limitations -- for instance, TNG300 is still small compared to the scale of the BAO for precision cosmology, and lacks statistics for the most massive halos at $\sim 10^{15}$ M$_\odot$, while TNG50 is still too low-resolution to resolve ultra-faint dwarf galaxies with $M_\star \lesssim 10^5$ M$_\odot$, globular clusters, or small-scale galactic features such as nuclear star clusters. We provide an overview and comparison between the specifications of all the TNG runs in \mbox{Table \ref{table_sims}}.

\begin{figure}[htb!]
\centerline{\includegraphics[angle=0,width=3.35in]{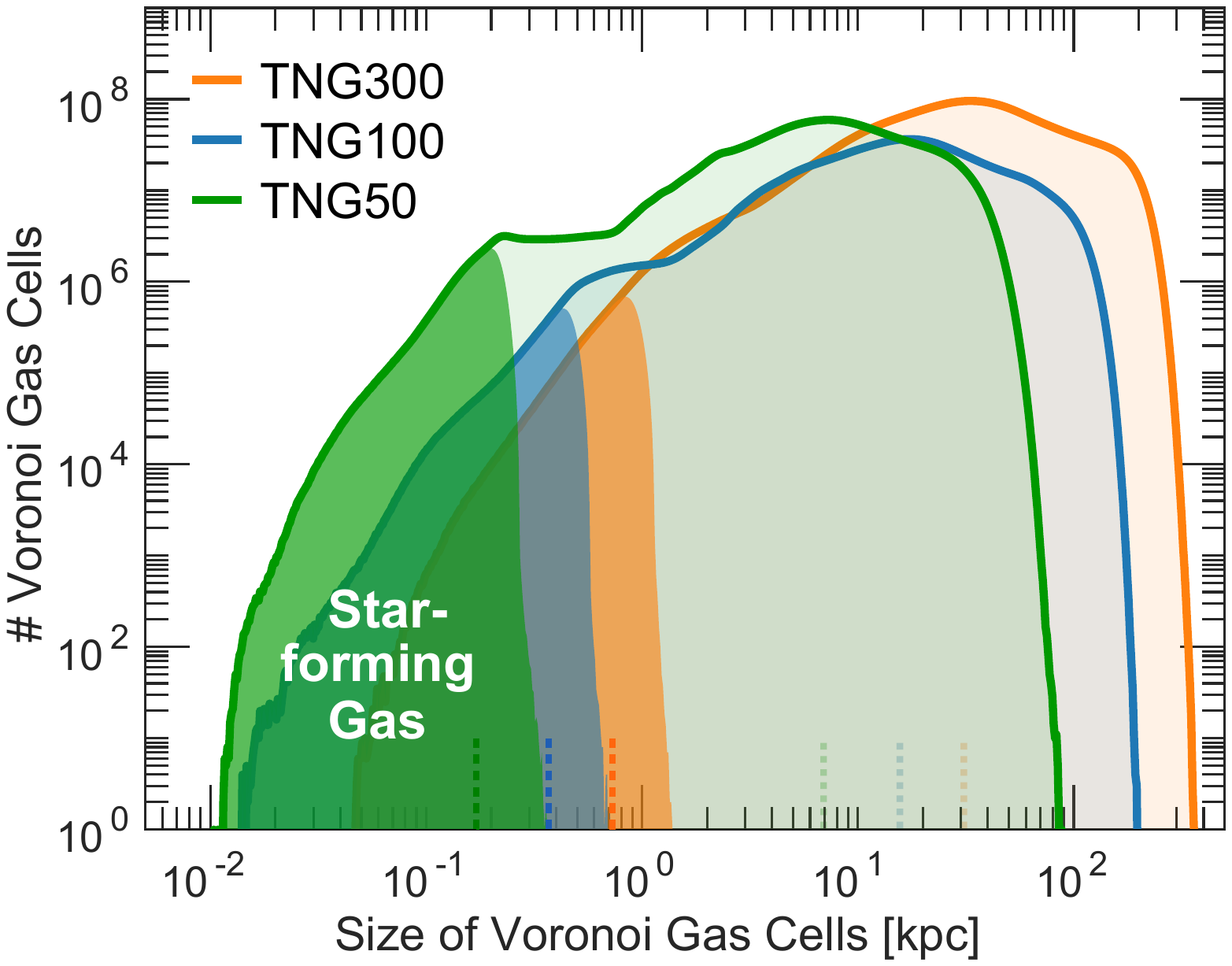}}
\caption{ Spatial resolution of the three high-resolution TNG simulations at $z\sim0$. The dark regions of the distributions highlight star-forming gas inside galaxies, the corresponding median values marked by dark vertical dotted lines.
 \label{fig_res}} 
\end{figure}

This data release includes the TNG50, TNG100 and TNG300 simulations in full. For each, snapshots at all 100 available redshifts, halo and subhalo catalogs at each snapshot, and two distinct merger trees are released. This includes three resolution levels of the 100 and 300 Mpc volumes, and four resolution levels of the 50 Mpc volume, decreasing in steps of eight in mass resolution (two in spatial resolution) across levels. The highest resolution realizations, TNG50-1, TNG100-1 and TNG300-1, include $2 \times 2160^3$, $2 \times 1820^3$ and $2 \times 2500^3$ resolution elements, respectively (see Table \ref{table_sims}). As the actual spatial resolution of cosmological hydrodynamical simulations is highly adaptive, it is poorly captured by a single number. Figure \ref{fig_res} therefore shows the distribution of Voronoi gas cell sizes in these three simulations, highlighting the high spatial resolution in star-forming gas -- i.e., within galaxies themselves. In contrast, the largest gas cells occur in the low-density intergalactic medium. 

All ten of the baryonic runs invoke, without modification and invariant across box and resolution, the fiducial ``full'' galaxy formation physics model of TNG. All ten runs are accompanied by matched, dark matter only (i.e. gravity-only) analogs. In addition, there are multiple, high time-resolution ``subboxes'', with up to 8000 snapshots each and time spacing down to one million years.

This paper serves as the data release for IllustrisTNG as a whole.

{\renewcommand{\arraystretch}{1.2}
\begin{table*}[!ht]
  \caption{Table of physical and numerical parameters for each of the resolution levels of the three flagship TNG simulations. The physical parameters are: the box volume, the box side-length, the initial number of gas cells, dark matter particles, and Monte Carlo tracer particles. The target baryon mass, the dark matter particle mass, the $z$\,=\,0 Plummer equivalent gravitational softening of the collisionless component, the same value in comoving units, and the minimum comoving value of the adaptive gas gravitational softenings. Additional characterizations of the gas resolution, measured at redshift zero: the minimum physical gas cell radius, the median gas cell radius, the mean radius of SFR$>$0 gas cells, the mean hydrogen number density of star-forming gas cells, and the maximum hydrogen gas density.}
  \label{table_sims}
  \begin{center}
    \begin{tabular}{lllllllll}
     \hline\hline
 Run           & Volume         & $L_{\rm box}$  & $N_{\rm GAS,DM}$ & $N_{\rm TRACER}$ & $m_{\rm baryon}$    & $m_{\rm DM}$        & $m_{\rm baryon}$        & $m_{\rm DM}$ \\
               & [\,cMpc$^3$\,] & [\,cMpc/$h$\,] & -                & -                & [\,M$_\odot / h$\,] & [\,M$_\odot / h$\,] & [\,10$^6$\,M$_\odot$\,] & [\,10$^6$\,M$_\odot$\,] \vspace{0.2em}\\ \hline
 TNG50-1       & $51.7^3$       & 35             & $2160^3$         & $1\times2160^3$  & $5.7 \times 10^4$   & $3.1 \times 10^5$   & 0.08                    & 0.45  \\
 TNG50-2       & $51.7^3$       & 35             & $1080^3$         & $1\times1080^3$  & $4.6 \times 10^5$   & $2.5 \times 10^6$   & 0.68                    & 3.63  \\
 TNG50-3       & $51.7^3$       & 35             & $540^3$          & $1\times540^3$   & $3.7 \times 10^6$   & $2.0 \times 10^7$   & 5.4                     & 29.0  \\
 TNG50-4       & $51.7^3$       & 35             & $270^3$          & $1\times270^3$   & $2.9 \times 10^7$   & $1.6 \times 10^8$   & 43.4                    & 232   \\
 TNG100-1      & $106.5^3$      & 75             & $1820^3$         & $2\times1820^3$  & $9.4 \times 10^5$   & $5.1 \times 10^6$   & 1.4                     & 7.5   \\
 TNG100-2      & $106.5^3$      & 75             & $910^3$          & $2\times910^3$   & $7.6 \times 10^6$   & $4.0 \times 10^7$   & 11.2                    & 59.7  \\
 TNG100-3      & $106.5^3$      & 75             & $455^3$          & $2\times455^3$   & $6.0 \times 10^7$   & $3.2 \times 10^8$   & 89.2                    & 478   \\
 TNG300-1      & $302.6^3$      & 205            & $2500^3$         & $1\times2500^3$  & $7.6 \times 10^6$   & $4.0 \times 10^7$   & 11                      & 59    \\
 TNG300-2      & $302.6^3$      & 205            & $1250^3$         & $1\times1250^3$  & $5.9 \times 10^7$   & $3.2 \times 10^8$   & 88                      & 470   \\
 TNG300-3      & $302.6^3$      & 205            & $625^3$          & $1\times625^3$   & $4.8 \times 10^8$   & $2.5 \times 10^9$   & 703                     & 3760  \\ \hline 
 TNG50-1-Dark  & $51.7^3$       & 35             & $2160^3$         & -                & -                   & $3.7 \times 10^5$   & -                       & 0.55  \\
 TNG50-2-Dark  & $51.7^3$       & 35             & $1080^3$         & -                & -                   & $2.9 \times 10^6$   & -                       & 4.31  \\
 TNG50-3-Dark  & $51.7^3$       & 35             & $540^3$          & -                & -                   & $2.3 \times 10^7$   & -                       & 34.5  \\
 TNG50-4-Dark  & $51.7^3$       & 35             & $270^3$          & -                & -                   & $1.9 \times 10^8$   & -                       & 275   \\
 TNG100-1-Dark & $106.5^3$      & 75             & $1820^3$         & -                & -                   & $6.0 \times 10^6$   & -                       & 8.9   \\
 TNG100-2-Dark & $106.5^3$      & 75             & $910^3$          & -                & -                   & $4.8 \times 10^7$   & -                       & 70.1  \\
 TNG100-3-Dark & $106.5^3$      & 75             & $455^3$          & -                & -                   & $3.8 \times 10^8$   & -                       & 567   \\
 TNG300-1-Dark & $302.6^3$      & 205            & $2500^3$         & -                & -                   & $7.0 \times 10^7$   & -                       & 47    \\
 TNG300-2-Dark & $302.6^3$      & 205            & $1250^3$         & -                & -                   & $3.8 \times 10^8$   & -                       & 588   \\
 TNG300-3-Dark & $302.6^3$      & 205            & $625^3$          & -                & -                   & $3.0 \times 10^9$   & -                       & 4470  \vspace{1em} \\ 

 \hline \hline
 Run      & $\epsilon_{\rm DM,\star}^{z=0}$ & $\epsilon_{\rm DM,\star}$ & $\epsilon_{\rm gas,min}$ & $r_{\rm cell,min}$ & $\bar{r}_{\rm cell}$ & $\bar{r}_{\rm cell,SF}$ & $\bar{n}_{\rm H,SF}$ & $n_{\rm H,max}$ \\
          & [\,kpc\,]                       & [\,ckpc/$h$\,]            & [\,ckpc/$h$\,]           & [\,pc\,]           & [\,kpc\,]            & [\,pc\,]                & [\,cm$^{-3}$\,]      & [\,cm$^{-3}$\,] \vspace{0.2em}\\ \hline
 TNG50-1  & 0.29                            & 0.39 $\rightarrow$ 0.195  & 0.05                     & 8                  & 5.8                  & 138                     & 0.8           & 650 \\ 
 TNG50-2  & 0.58                            & 0.78 $\rightarrow$ 0.39   & 0.1                      & 19                 & 12.9                 & 282                     & 0.7                  & 620  \\
 TNG50-3  & 1.15                            & 1.56 $\rightarrow$ 0.78   & 0.2                      & 65                 & 25.0                 & 562                     & 0.6                  & 80   \\
 TNG50-4  & 2.30                            & 3.12 $\rightarrow$ 1.56   & 0.4                      & 170                & 50.1                 & 1080                    & 0.5                  & 35   \\
 TNG100-1 & 0.74                            & 1.0 $\rightarrow$ 0.5     & 0.125                    & 14                 & 15.8                 & 355                     & 1.0                  & 3040 \\
 TNG100-2 & 1.48                            & 2.0 $\rightarrow$ 1.0     & 0.25                     & 74                 & 31.2                 & 720                     & 0.6                  & 185  \\
 TNG100-3 & 2.95                            & 4.0 $\rightarrow$ 2.0     & 0.5                      & 260                & 63.8                 & 1410                    & 0.5                  & 30   \\
 TNG300-1 & 1.48                            & 2.0 $\rightarrow$ 1.0     & 0.25                     & 47                 & 31.2                 & 715                     & 0.6                  & 490  \\
 TNG300-2 & 2.95                            & 4.0 $\rightarrow$ 2.0     & 0.5                      & 120                & 63.8                 & 1420                    & 0.5                  & 235  \\
 TNG300-3 & 5.90                            & 8.0 $\rightarrow$ 4.0     & 1.0                      & 519                & 153                  & 3070                    & 0.4                  & 30   \\

 \hline
    \end{tabular}
  \end{center}
\end{table*}}

\subsection{Physical Models and Numerical Methods}

All of the TNG runs start from cosmologically motivated initial conditions, assuming a cosmology consistent with the \cite{planck2015_xiii} results ($\Omega_{\Lambda,0}=0.6911$, $\Omega_{m,0}=0.3089$, $\Omega_{b,0}=0.0486$, $\sigma_8=0.8159$, $n_s=0.9667$ and $h=0.6774$), with Newtonian self-gravity solved in an expanding Universe. All of the baryonic TNG runs include the following additional physical components:
(1) Primordial and metal-line radiative cooling in the presence of an ionizing background radiation field which is redshift-dependent and spatially uniform, with additional self-shielding corrections.
(2) Stochastic star formation in dense ISM gas above a threshold density criterion.
(3) Pressurization of the ISM due to unresolved supernovae using an effective equation of state model for the two-phase medium.
(4) Evolution of stellar populations, with associated chemical enrichment and mass loss (gas recycling), accounting for SN Ia/II, AGB stars, and NS-NS mergers.
(5) Stellar feedback: galactic-scale outflows with an energy-driven, kinetic wind scheme.
(6) Seeding and growth of supermassive blackholes.
(7) Supermassive blackhole feedback: accreting BHs release energy in two modes, at high-accretion rates (`quasar' mode) and low-accretion rates (`kinetic wind' mode). Radiative proximity effects from AGN affect nearby gas cooling.
(8) Magnetic fields: amplification of a small, primordial seed field and dynamical impact under the assumption of ideal MHD.

{\renewcommand{\arraystretch}{1.2}
\begin{table*}
  \caption{Comparison of key model changes between Illustris and IllustrisTNG. For full details and a more comprehensive comparison including numerical parameter differences, see Table 1 of \protect\cite{pillepich18a} and the two TNG methods papers in general.}
  \label{table_model}
  \begin{center}
    \begin{tabular}{lcc}
     \hline\hline
Simulation Aspect & Illustris & TNG (50/100/300) \\
 \hline 
Magnetic Fields & no & ideal MHD \citep{pakmor11} \\
BH Low-State Feedback & `Radio' Bubbles & BH-driven wind (kinetic kick) \\
BH Accretion & Boosted Bondi-Hoyle ($\alpha=100$) & Un-boosted Bondi-Hoyle \\
BH Seed mass & $10^5$ M$_\odot$/h & $8 \times 10^5$ M$_\odot$/h \\
Winds (Directionality) & bi-polar ($\vec{v}_{\rm gas} \times \nabla \phi_{\rm grav}$) & isotropic \\
Winds (Thermal Content) & cold & warm (10\%) \\
Winds (Velocity) & $\propto \sigma_{\rm DM}$ & + scaling with H(z), and $v_{\rm min}$ \\
Winds (Energy) & constant per unit SFR & + metallicity dependence in $\eta$ \\
Stellar Evolution & Illustris Yields & TNG Yields \\
Metals Tagging & - & SNIa, SNII, AGB, NSNS, FeSNIa, FeSNII \\
Shock Finder & no & yes \citep{schaal15} \\
 \hline
    \end{tabular}
  \end{center}
\end{table*}}

For complete details on the behavior, implementation, parameter selection, and validation of these physical models, see the two TNG methods papers: \cite{weinberger17} and \cite{pillepich18a}. Table \ref{table_model} provides an abridged list of the key differences between Illustris and TNG. We note that the TNG model has been designed (i.e. `calibrated', or `tuned') to broadly reproduce several basic, observed galaxy properties and statistics. These are: the galaxy stellar mass function and the stellar-to-halo mass relation, the total gas mass content within the virial radius ($r_{500}$) of massive groups, the stellar mass -- stellar size and the BH--galaxy mass relations all at $z=0$, in addition to the overall shape of the cosmic star formation rate density at $z\lesssim10$ \citep[see][for a discussion]{pillepich18a}.

The TNG simulations use the moving-mesh {\sc Arepo} code \citep{spr10} which solves the equations of continuum magnetohydrodynamics \citep[MHD;][]{pakmor11,pakmor13} coupled with self-gravity. The latter is computed with the Tree-PM approach, while the fluid dynamics employs a Godunov (finite-volume) type method, with a spatial discretization based on an unstructured, moving, Voronoi tessellation of the domain. The Voronoi mesh is generated from a set of control points which move with the local fluid velocity modulo mesh regularization corrections. Assuming ideal MHD, an 8-wave Powell cleaning scheme maintains the zero divergence constraint. The previous MUSCL-Hancock scheme has been replaced with a time integration approach following Heun's method, and the original Green-Gauss method for gradient estimation of primitive fluid quantities has been replaced with a least-squares method, obtaining second order convergence in the hydrodynamics \citep{pakmor16}. The long-range FFT calculation employs a new column-based MPI-parallel decomposition, while the gravity solver has been rewritten based on a recursive splitting of the N-body Hamiltonian into short- and long- timescale systems (as in {\sc Gadget-4}, \textcolor{blue}{Springel in prep.}). The code is second order in space, and with hierarchical adaptive time-stepping, also second order in time. Of order 10 million individual timesteps are required to evolve the high-resolution runs to redshift zero.

During the simulation we employ a Monte Carlo tracer particle scheme \citep{genel13} to follow the Lagrangian evolution of baryons. An on-the-fly cosmic shock finder is coupled to the code \citep{schaal15,schaal16}. Group catalogs are computed during the simulations using the {\sc FoF} and {\sc Subfind} \citep{spr01} substructure identification algorithms.

\subsection{Model Validation and Early Findings} \label{sec_earlyresults}

TNG has been shown to produce observationally consistent results in several regimes beyond those adopted to calibrate the model. Some examples regarding galaxy populations, galactic structural and stellar population properties include: the shapes and widths of the red sequence and blue cloud of SDSS galaxies \citep{nelson18a}; the shapes and normalizations of the galaxy stellar mass functions up to $z\sim4$ \citep{pillepich18b}; the spatial clustering of red vs. blue galaxies from tens of kpc to tens of Mpc separations \citep{springel18}; the spread in Europium abundance of metal-poor stars in Milky Way like halos \citep{naiman18}; the emergence of a population of quenched galaxies both at low \citep{weinberger18} and high redshift \citep{habouzit18}; stellar sizes up to $z\sim2$, including separate star-forming and quiescent populations \citep{genel18}; the $z=0$ and evolution of the gas-phase mass-metallicity relation \citep{torrey18}; the dark matter fractions within the extended bodies of massive galaxies at $z=0$ in comparison to e.g. SLUGGS results \citep{lovell18}; and the optical morphologies of galaxies in comparison to Pan-STARRS observations \citep{rodriguezgomez18}.

The IllustrisTNG model also reproduces a broad range of unusual galaxies, tracing tails of the galaxy population, including low surface brightness galaxies \citep{zhu18} and jellyfish, ram-pressure stripped galaxies \citep{yun18}. The large-volume of TNG300 helps demonstrate reasonable agreement in several galaxy cluster, intra-cluster and circumgalactic medium properties -- for example, the scaling relations between total radio power and X-ray luminosity, total mass, and Sunyaev-Zel'dovich parameter of massive haloes \citep{marinacci18}; the distribution of metals in the intra-cluster plasma \citep{vog18a}; the observed fraction of cool core clusters \citep{barnes18}; and the OVI content of the circumgalactic media around galaxies from surveys at low redshift including COS-Halos and eCGM \citep{nelson18b}.

IllustrisTNG is also producing novel insights on the formation and evolution of galaxies. For instance, halo mass alone is a good predictor for the entire stellar mass profile of massive galaxies \citep{pillepich18b}; the metal enrichment of cluster galaxies is higher than field counterparts at fixed mass and this enhancement is present pre-infall \citep{gupta18}; star-forming and quenched galaxies take distinct evolutionary pathways across the galaxy size-mass plane \citep{genel18} and exhibit systematically different column densities of OVI ions \citep{nelson18b} and different magnetic-field strengths \citep{nelson18a} at fixed galaxy stellar mass, as well as different magnetic-field topologies \citep{marinacci18}. Galaxies oscillate around the star formation main sequence and the mass-metallicity relations over similar timescales and often in an anti-correlated fashion \citep{torrey18b}; the presence of jellyfish galaxies is signaled by large-scale bow shocks in their surrounding intra-cluster medium \citep{yun18}; baryonic processes affect the matter power spectrum across a range of scales \citep{springel18} and steepen the inner power-law total density profiles of early-type galaxies \citep{wang18}; a significant number of OVII, OVIII \citep{nelson18b} and NeIX \citep{martizzi18} absorption systems are expected to be detectable by future X-ray telescopes like ATHENA. 

IllustrisTNG has also been used to generate mock 21-cm maps \citep{villaescusa18} and estimates of the molecular hydrogen budget \citep{diemer18} in central and satellite galaxies in the local \citep{stevens18} as well as in the high-redshift Universe as probed by ALMA \citep{popping19}. Finally, TNG provides a test bed to explore future observational applications of machine learning techniques: for example, the use of Deep Neural Networks to estimate galaxy cluster masses from Chandra X-ray mock images \citep{ntampaka18} or optical morphologies versus SDSS \citep{huertascompany19}.

See the up to date list of results\footnote{\url{www.tng-project.org/results}} for additional references. Please note that on this page we provide, and will continue to release, data files accompanying published papers as appropriate. For instance, electronic versions of tables, and data points from key lines and figures, to enable comparisons with other results. These are available with small {\sc [data]} links next to each paper.

\subsection{Breadth of Simulated Data}

\begin{figure*}[htb!]
\centerline{\includegraphics[angle=0,width=6.5in]{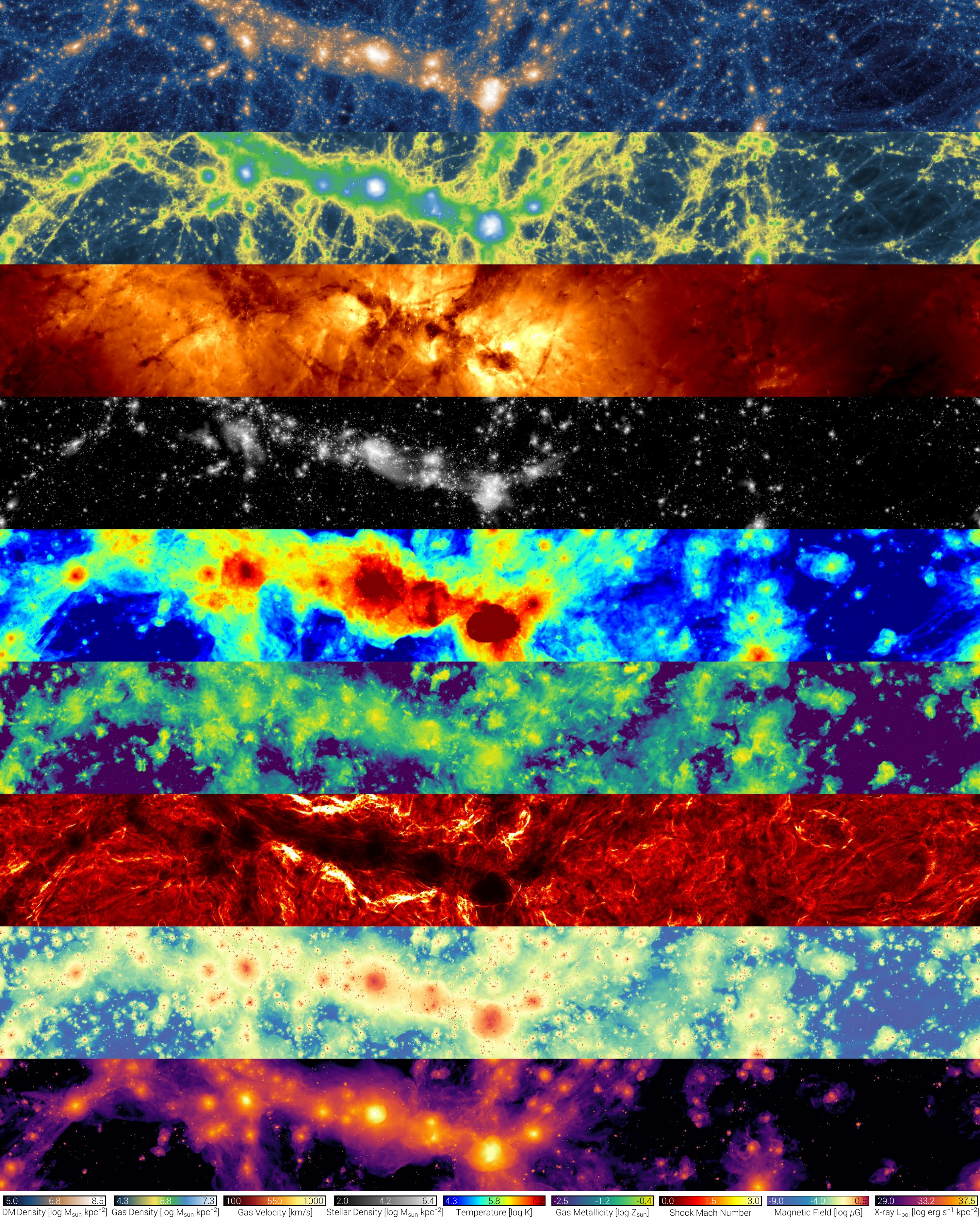}}
\caption{ Overview of the variety of physical information accessible in the different matter components of the TNG simulations. From top to bottom: dark matter density, gas density, gas velocity field, stellar mass density, gas temperature, gas-phase metallicity, shock mach number, magnetic field strength, and x-ray luminosity. Each panel shows the same $\sim 110 \times 14 \times 37$ Mpc volume of TNG100-1 at $z=0$.
 \label{fig_fields}} 
\end{figure*}

All of the observational validations and early results from TNG100 and TNG300 demonstrate the broad applications of the IllustrisTNG simulations. To give a sense of the expansive scope, the richness of the resulting data products, and the potential for wide applications across many areas of galaxy formation and cosmology, Figure \ref{fig_fields} visualizes the TNG100 simulation at redshift zero. Each slice reveals a view into the synthetic IllustrisTNG universe. Together, they range from purely theoretical quantities to directly observable signatures, spanning across the baryonic and non-baryonic matter components of the simulation: dark matter, gas, stars, and blackholes.

The wealth of available information in the simulation outputs translates directly into the wide range of astrophysical phenomena which can be explored with the TNG simulations.


\section{Data Products} \label{sDataProducts}

We release all 100 snapshots of the IllustrisTNG cosmological volumes. These include up to five types of resolution elements (dark matter particles, gas cells, gas tracers, stellar and stellar wind particles, and supermassive blackholes). The same volumes are available at multiple resolutions: high (-1 suffix, e.g. TNG100-1), intermediate (-2 suffix), and low (-3 suffix), always separated by a factor of two (eight) in spatial (mass) resolution. At each resolution, these `baryonic' runs include the fiducial TNG model for galaxy formation physics. Each baryonic run is matched to its dark matter only analog (-Dark suffix).

For all runs, at every snapshot, two types of group catalogs are provided: friends-of-friends (FoF) halo catalogs, and {\sc Subfind} subhalo catalogs. In postprocessing, these catalogs are used to generate two distinct merger trees, which are both released: {\sc SubLink}, and {\sc LHaloTree}. Finally, supplementary data catalogs containing additional computations and modeling, and focusing on a variety of topics, are being continually created and released. All these data types are described below.

\subsection{Snapshots}

\subsubsection{Snapshot Organization}

There are 100 snapshots stored for every run. These include all particles/cells in the whole volume. The complete snapshot listings, spacings and redshifts can be found online. Note that, unlike in Illustris, TNG contains two different types of snapshots: `full' and `mini'. While both encompass the entire volume, `mini' snapshots only have a subset of particle fields available (detailed online). In TNG, twenty snapshots are full, while the remaining 80 are mini. The 20 full snapshots are given in Table \ref{table_snaps}. Every snapshot is stored on-disk in a series of `chunks', which are more manageable, smaller HDF5 files -- additional details are provided in Table \ref{table_chunks} of the appendix.

\begin{table}[htb!]
\footnotesize
  \caption{Abridged snapshot list for TNG runs: snapshot number together with the corresponding scalefactor and redshift. The twenty snapshots shown here are the `full' snapshots, while the remaining eighty are `mini' snapshots with a subset of fields.}
  \label{table_snaps}
  \begin{center}
\renewcommand{\arraystretch}{1.3}
    \begin{tabular}{cllcll}
    \hline
   Snap & $a$ & $z$ & Snap & $a$ & $z$ \\ \hline\hline
   2  & 0.0769 & 12  & 33 & 0.3333 & 2   \\
   3  & 0.0833 & 11  & 40 & 0.4    & 1.5 \\
   4  & 0.0909 & 10  & 50 & 0.5    & 1   \\
   6  & 0.1    & 9   & 59 & 0.5882 & 0.7 \\
   8  & 0.1111 & 8   & 67 & 0.6667 & 0.5 \\
   11 & 0.125  & 7   & 72 & 0.7143 & 0.4 \\
   13 & 0.1429 & 6   & 78 & 0.7692 & 0.3 \\
   17 & 0.1667 & 5   & 84 & 0.8333 & 0.2 \\
   21 & 0.2    & 4   & 91 & 0.9091 & 0.1 \\
   25 & 0.25   & 3   & 99 & 1      & 0   \\ \hline
    \end{tabular}
  \end{center}
\end{table}

\begin{figure}[tb!]
\centerline{\includegraphics[angle=0,width=3.3in]{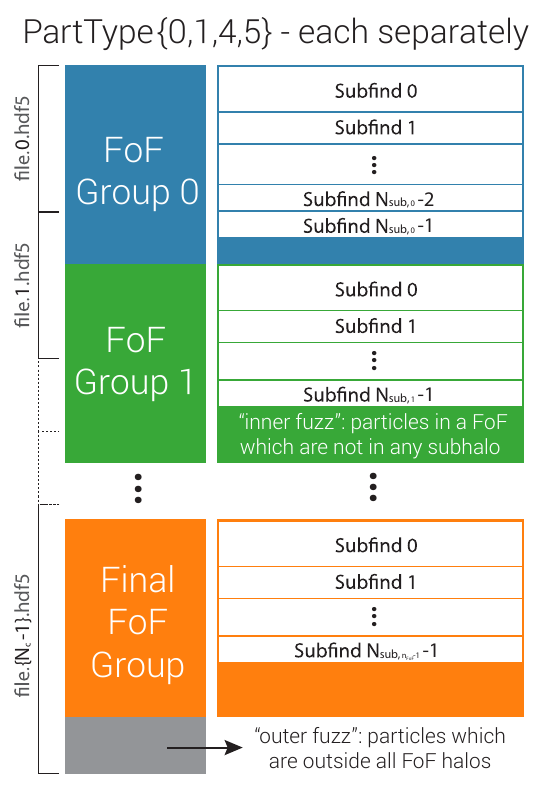}}
\caption{ Illustration of the organization of particle/cell data within a snapshot for one particle type (e.g dark matter). Therein, particle order is set by a global sort of the following fields in this order: FoF group number, {\sc Subfind} subhalo number, binding energy. As a result, FOF halos are contiguous, although they can span file chunks. {\sc Subfind} subhalos are only contiguous within a single group, being separated between groups by an ``inner fuzz'' of all FOF particles not bound to any subhalo. ``Outer fuzz'' particles outside all halos are placed at the end of each snapshot.
 \label{fig_snap_schematic}} 
\end{figure}

Note that, just as in Illustris, the snapshot data is not organized according to spatial position. Rather, particles within a snapshot are sorted based on their group/subgroup memberships, according to the {\sc FoF} or {\sc Subfind} algorithms. Within each particle type, the sort order is: group number, subgroup number, and then binding energy. Particles/cells belonging to the group but not to any of its subhalos (``inner fuzz'') are included after the last subhalo of each group. In Figure \ref{fig_snap_schematic} we show a schematic of the particle organization \citep[as in][]{nelson15b}, for \textit{one particle type}. Note that halos may happen to be stored across multiple, subsequent file chunks, and different particle types of a halo are in general stored in different sets of file chunks.

\subsubsection{Snapshot Contents}

Each HDF5 snapshot contains several groups: `Header', `Parameters', `Configuration', and five additional `PartTypeX' groups, for the following particle types (DM only runs have a single PartType1 group):\par\nobreak

\vspace{0.5em}
\begin{itemize}[leftmargin=2em]
\setlength\itemsep{0.2em}
\item PartType0 - GAS
\item PartType1 - DM
\item PartType2 - (unused)
\item PartType3 - TRACERS
\item PartType4 - STARS \& WIND PARTICLES
\item PartType5 - BLACK HOLES
\end{itemize}
\vspace{0.5em}

The `Header' group contains a number of attributes giving metadata about the simulation and snapshot. The `Parameters' and `Configuration' groups provide the complete set of run-time parameter and compile-time configuration options used to run TNG. That is, they encode the fiducial ``TNG Galaxy Formation Model''. Many will clearly map to Table 1 of \cite{pillepich18a}, while others deal with more numerical/technical options. In the future, together with the release of the TNG initial conditions and the TNG code base, this will enable any of the TNG simulations to be reproduced.

The complete snapshot field listings of the `PartTypeX' groups, including dimensions, units and descriptions, are given online. The general system of units is ${\rm kpc}/h$ for lengths, $10^{10} {\rm M}_\odot/h$ for masses, and ${\rm km/s}$ for velocities. Comoving quantities can be converted to the corresponding physical ones by multiplying by the appropriate power of the scale factor $a$. New fields in TNG, not previously available in the original Illustris, are specially highlighted.

With respect to Illustris, the following new fields are generally available in the snapshots: (i) EnergyDissipation and Machnumber, giving the output of the on-the-fly shock finder, (ii) GFM\_Metals, giving the individual element abundances of the nine tracked species (H, He, C, N, O, Ne, Mg, Si, Fe), (iii) GFM\_MetalsTagged, metal tracking as described below, (iv) MagneticField and MagneticFieldDivergence, providing the primary result of the MHD solver.

\subsubsection{Tagged Metals}

\begin{table*}[tb!]
\footnotesize
  \caption{Configuration of each of the subboxes for all three TNG volumes, including position and size within the periodic parent simulation, and description of the environment contained within.}
  \label{table_subbox1}
  \begin{center}
\renewcommand{\arraystretch}{1.5}
    \begin{tabular}{cccccc}
    \hline
  Subbox & Environment & Center Position [Code Units] & Box Size & $f_{\rm vol}$ [\%] \\ \hline\hline
  TNG100 Subbox-0 & Crowded, including a $5 \times 10^{13} {\rm M}_\odot$ halo & (9000, 17000, 63000)  & 7.5 ${\rm cMpc}/h$ & 0.1 \\
  TNG100 Subbox-1 & Less crowded, including several $> 10^{12} {\rm M}_\odot$ halos  & (37000, 43500, 67500) & 7.5 ${\rm cMpc}/h$ & 0.1 \vspace{0.5em}\\

  TNG300 Subbox-0 & Massive cluster ($\sim2 \times 10^{15} M_{\odot}$) merging at z=0 & (44, 49, 148) * 1000 & 15 ${\rm cMpc}/h$ & 0.04 \\
  TNG300 Subbox-1 & Crowded, above average \# of halos above $10^{13} M_{\odot}$ & (20, 175, 15) * 1000 & 15 ${\rm cMpc}/h$ & 0.04 \\
  TNG300 Subbox-2 & Semi-underdense, one local group analog at z=0 & (169, 97.9, 138) * 1000 & 10 ${\rm cMpc}/h$ & 0.01 \vspace{0.5em}\\
  
  TNG50 Subbox-0  & Somewhat-crowded ($\sim$6 MWs) &  (26000, 10000, 26500)  & 4.0 ${\rm cMpc}/h$ & 0.15 \\
  TNG50 Subbox-1  & Low-density, many dwarfs, no halos $>5 \times 10^{10}$ M$_{\odot}$ & (12500, 10000, 22500)  & 4.0 ${\rm cMpc}/h$ & 0.15 \\
  TNG50 Subbox-2  & Most massive cluster ($2 \times 10^{14}$ M$_{\odot}$ at z=0) & (7300, 24500, 21500)  & 5.0 ${\rm cMpc}/h$ & 0.3 \\ \hline
    \end{tabular}
  \end{center}
\end{table*}

\begin{table}[tb!]
\footnotesize
  \caption{Details of the subbox snapshots: the number and approximate time resolution $\Delta t$ at three redshifts: $z=6$, $z=2$, and $z=0$. Every subbox for a given volume and resolution combination has the same output times.}
  \label{table_subbox2}
  \begin{center}
\renewcommand{\arraystretch}{1.5}
    \begin{tabular}{crrrrr}
    \hline
  Run & $N_{\rm snap}$ & $\Delta t_{(z=6)}$ & $\Delta t_{(z=2)}$ & $\Delta t_{(z=0)}$ \\ \hline\hline
  TNG100-3 & 2431  &  $\sim$4 Myr   &  $\sim$7 Myr   &  $\sim$19 Myr \\
  TNG100-2 & 4380  &  $\sim$2 Myr   &  $\sim$4 Myr   &  $\sim$10 Myr \\
  TNG100-1 & 7908  &  $\sim$1 Myr   &  $\sim$1.5 Myr &  $\sim$2.5 Myr\\
  TNG300-3 & 2050  &  $\sim$6 Myr   &  $\sim$11 Myr  &  $\sim$8 Myr \\
  TNG300-2 & 3045  &  $\sim$3 Myr   &  $\sim$6 Myr   &  $\sim$4 Myr \\
  TNG300-1 & 2449  &  $\sim$1.5 Myr &  $\sim$4 Myr   &  $\sim$6 Myr \\
  TNG50-4  & 2333  &  $\sim$7 Myr   &  $\sim$6 Myr   &  $\sim$8 Myr \\
  TNG50-3  & 4006  &  $\sim$2 Myr   &  $\sim$3 Myr   &  $\sim$4 Myr \\
  TNG50-2  & 1895  &  $\sim$3 Myr   &  $\sim$6 Myr   &  $\sim$8 Myr \\
  TNG50-1  & $\sim$3600 &  $\sim$3 Myr   &  $\sim$2 Myr   &  $\sim$2 Myr \\ \hline
    \end{tabular}
  \end{center}
\end{table}

The units of all the entries of GFM\_MetalsTagged field, except for NSNS, are the same as GFM\_Metals: dimensionless mass ratios. Summing all elements of GFM\_Metals heavier than Helium recovers the sum of the three tags SNIa+SNII+AGB. Likewise, the Fe entry of GFM\_Metals roughly equals the sum of FeSNIa+FeSNII, modulo the small amount of iron consumed (i.e. negative contribution) by AGB winds. The fields are (in order):

\vspace{0.5em}
\begin{itemize}[leftmargin=2em]
\setlength\itemsep{0.2em}
\item SNIa (0): The total metals ejected by Type Ia SN.

\item SNII (1): The total metals ejected by Type II SN.

\item AGB (2): the total metals ejected by stellar winds, which is dominated by AGB stars.

\item NSNS (3): the total mass ejected from NS-NS merger events, which are modeled stochastically (i.e. no fractional events) with a DTD scheme similar to that used for SNIa, except with a different $\tau$ value. Note that the units of NSNS are arbitrary. To obtain physical values in units of solar masses, this field must be multiplied by $\alpha / \alpha_0$ where $\alpha$ is the desired mass ejected per NS-NS merger event, and $\alpha_0$ is the base value (arbitrary) used in the simulation, e.g. \cite{shen15} take $\alpha = 0.05 \rm{M}_{\odot}$. The value of $\alpha_0$ varies by run, and it is 0.05 for all TNG100 runs, and 5000.0 for all TNG300 and TNG50 runs. See \cite{naiman18} for more details.

\item FeSNIa (4): The total iron ejected by Type Ia SN.

\item FeSNII (5): The total iron ejected by Type II SN.
\end{itemize}
\vspace{0.5em}

Note a somewhat subtle but fundamental detail: these tags do not isolate where a given heavy element was created, but rather identify the last star it was ejected from. This can be problematic since, for example, AGB winds create little iron, but eject a significant amount of iron which was previously created by SnIa and SNII at earlier epochs. The FeSNIa field is, for example, more accurately described as `the total iron ejected by type Ia supernovae not yet consumed and re-ejected from another star'.

\subsubsection{Subboxes}

Separate `subbox' cutouts exist for each baryonic run. These are spatial cutouts of fixed comoving size and fixed comoving coordinates, and the primary benefit is that their time resolution is significantly better than that of the main snapshots -- details are provided in Tables \ref{table_subbox1} and \ref{table_subbox2}. These snapshots are useful for some types of analysis and science questions requiring high time-resolution data, and for creating time-evolving visualizations. There are two subboxes for TNG100 (corresponding to the original Illustris subboxes \#0 and \#2, the latter increased in size), and three subboxes for TNG50 and TNG300. Note that subboxes, unlike full boxes, are not periodic.

The subboxes sample different areas of the large boxes, roughly described by the environment column in Table \ref{table_subbox1}. The particle fields are all identical to the main snapshots, except that the particles/cells are not sorted by their group membership, since no group catalogs exist for subbox snapshots.

\subsection{Group Catalogs}

Group catalogs give the results of substructure identification, and broadly contain two types of objects: dark matter halos (either {\sc FoF} halos or central subhalos) and galaxies themselves (the inner stellar component of subhalos, either centrals or satellites). There is one group catalog produced for each snapshot, which includes both {\sc FoF} and {\sc Subfind} objects. The group files are split into a small number of sub-files, just as with the raw snapshots. In TNG, these files are called {\sc fof\_subhalo\_tab\_*}, whereas in original Illustris they were called {\sc groups\_*} (they are otherwise essentially identical). Every HDF5 group catalog contains the following groups: Header, Group, and Subhalo. The IDs of the member particles of each group/subgroup are not stored in the group catalog files. Instead, particles/cells in the snapshot files are ordered according to group membership. 

In order to reduce confusion, we adopt the following terminology when referring to different types of objects.
``Group'', ``FoF Group'', and ``FoF Halo'' all refer to halos. ``Subgroup'', ``Subhalo'', and ``Subfind Group'' all refer to subhalos.
The first (most massive) subgroup of each halo is the ``Primary Subgroup'' or ``Central Subgroup''.
All other following subgroups within the same halo are ``Secondary Subgroups'', or ``Satellite Subgroups''.

\textbf{FoF Groups.} The Group fields are derived with a standard friends-of-friends (FoF) algorithm with linking length $b=0.2$. The FoF algorithm is run on the dark matter particles, and the other types (gas, stars, BHs) are attached to the same groups as their nearest DM particle. 
\textbf{Subfind Groups.} The Subhalo fields are derived with the {\sc Subfind} algorithm. In identifying gravitationally bound substructures the method considers all particle types and assigns them to subhalos as appropriate. 

Complete documentation for the TNG group catalogs, comprising FoF halos as well as Subfind subhalos, is available online. Differences and additions with respect to original Illustris are highlighted.

\subsection{Merger Trees}

Merger trees have been created for the TNG simulations using {\sc SubLink} \citep{rodriguezgomez15} and {\sc LHaloTree} \citep{spr05}. In the population average sense the different merger trees give similar results. In more detail, the exact merger history or mass assembly history for any given halo may differ. For a particular science goal, one type of tree may be more or less useful, and users are free to use whichever they prefer. We generally recommend use of the {\sc SubLink} trees as a first option, as they are more efficiently stored and accessible. 

Trees can be `walked', i.e. the descendants or progenitors of a given subhalo can be determined, thus linking objects across snapshots saved at different points in time. Main branches, such as the main progenitor branch (MPB), as well as full trees can be extracted. Examples of walking the tree are provided in the example scripts. For the technical details, algorithmic descriptions, and storage structures of the trees, please refer to \cite{nelson15b} and the online documentation -- we omit these details here.

\subsubsection{SubLink}

The {\sc SubLink} merger tree is one large data structure split across several sequential HDF5 files named \linebreak{\sc tree\_extended.[fileNum].hdf5}, where {\sc [fileNum]} goes from e.g. 0 to 19 for the TNG100-1 run, and 0 to 125 for the TNG300-1 run.

\subsubsection{LHaloTree}

The {\sc LHaloTree} merger tree is one large data structure split across several HDF5 files named \linebreak{\sc trees\_sf1\_99.[chunkNum].hdf5}, where TNG100-1 has for instance 80 chunks enumerated by {\sc [chunkNum]}, while TNG300-1 has 320. Within each file there are a number of HDF5 groups named ``TreeX'', each of which represents one disjoint merger tree.

\subsubsection{Offsets Files}

As described above, snapshot particle data is ordered by the subhalo each particle belongs to. To facilitate rapid loading of snapshot data, particle `offset' numbers provide the location where particles belonging to each subhalo begin. Most simply, offsets describe where in the group catalog files to find a specific halo/subhalo, and where in the snapshot files to find the start of the particles of a given halo/subhalo.

To use the helper scripts (provided online) for working with the actual data files (snapshots or group catalogs) on a local machine, then it \textbf{is required} to download the offset file(s) for the snapshot(s) of interest. The offsets are \textbf{not required} when using the web-based API or analyzing the particle cutouts it provides, for instance.

Note that in Illustris, offsets were embedded inside the group catalog files for convenience. In TNG however, we have kept offsets as separate files called {\sc offsets\_*.hdf5} (one per snapshot), which must be downloaded as well.

\subsubsection{The `simulation.hdf5' file}

Each run has a single file called `simulation.hdf5' which is purely optional, for convenience, and not required by any of the public scripts. Its purpose is to encapsulate all data of an entire simulation into a single file.

To accomplish this, we make advantage of a new feature of the HDF5 library called ``virtual datasets''. A virtual dataset is a collection of symbolic links to one or more datasets in other HDF5 file(s), where these symlinks can refer to subsets of a dataset, in either the source or target of the link. The simulation.hdf5 is thus a large collection of links, which refer to other files which actually contain data. In order to use it, the corresponding files must also be downloaded (e.g. of snapshots, group catalogs, or supplementary data catalogs).

Using this resource, the division of snapshots and group catalogs over multiple file chunks is no longer relevant. Loading particle data from snapshots or subhalo or halo fields from group catalogs become one line operations. It also makes loading the particles of a given halo or subhalo using the offset information trivial. 
Finally, supplementary data catalogs (either those we provide, or similar user-run computations) can be `virtually' inserted as datasets in snapshots or group catalogs. This provides a clean way to organize post-processing computations which produce additional values for halos, subhalos, or individual particles/cells. Such data can then be loaded with the same scripts (and same syntax) as `original' snapshot/group catalog fields.

We refer to the online documentation for examples of each use case as well as technical requirements, namely a relatively new version (1.10+) of the HDF5 library.

\subsection{Initial Conditions}

We provide as part of this release the initial conditions for all TNG volumes as well as the original Illustris volumes. These were created with the Zeldovich approximation and the {\sc N-GenIC} code \citep{springel15}. Each particular realization was chosen from among tends of random realizations of the same volume as the most average, based on sinspection of the $z=0$ power spectrum and/or dark matter halo mass function -- see \cite{vog14a} and \textcolor{blue}{Pillepich et al. (in prep)} for details. Each IC is a single HDF5 file with self-evident structure: the coordinates, velocities, and IDs of the set of total matter particles at $z=127$, the starting redshift for all runs. These ICs were used as is for dark-matter only simulations, while for baryonic runs total matter particles were split upon initialization in the usual way, into dark matter and gas, according to the cosmic baryon fraction and offsetting in space by half the mean interparticle spacing. These ICs can be run by e.g. {\sc Gadget} or {\sc Arepo} as is, or easily converted into other data formats.

\subsection{Supplementary Data Catalogs}

Many additional data products have been computed in post-processing, based on the raw simulation outputs. These are typically in support of specific projects and analysis in a published paper, after which the author makes the underlying data catalog public. Many such catalogs have been made available for the original Illustris simulation, and the majority of these will also be recalculated for TNG. We provide a list of TNG supplementary data catalogs which are now available or which we anticipate to release in the near future:

\begin{enumerate}[label=(\Alph*)]
  \item Tracer Tracks -- time-evolution of Monte Carlo tracer properties for TNG100 (\textcolor{blue}{Nelson et al. in prep}).
  \item Stellar Mass, Star Formation Rates -- multi-aperture and resolution corrected masses, time-averaged SFRs \citep{pillepich18b}.
  \item Stellar Circularities, Angular Momenta, and Axis Ratios -- for the stellar components of galaxies, as for Illustris \citep{genel15}.
  \item Subhalo Matching Between Runs -- cross-matching subhalos between baryonic and dark-matter only runs, between runs at different resolutions, and between TNG100 and Illustris \citep{lovell18,nelson15b,rodriguezgomez15,rodriguezgomez17}.
  \item Stellar Projected Sizes -- half-light radii of TNG100 galaxies \citep{genel18}.
  \item Blackhole Mergers and Details -- records of BH-BH mergers and high time-resolution BH details, as for Illustris \citep{kelley17,blecha16}, and with an updated approach (\textcolor{blue}{Katz et al. in prep}).
  \item Stellar Assembly -- in-situ versus ex-situ stellar growth, as for Illustris \citep{rodriguezgomez16,rodriguezgomez17}.
  \item Subbox Subhalo List -- record of which subhalos exist in what subboxes across particular redshift ranges, and interpolated properties \citep{nelson19b}
  \item Molecular and Atomic Hydrogen (HI+H2) -- decomposition of the neutral hydrogen in gas cells and galaxies into HI/H2 masses \citep{diemer18,stevens18}.
  \item Halo/galaxy angular momentum and baryon content -- measurements of spherical overdensity values, as for Illustris \citep{zjupa17}.
  \item SDSS Photometry and Mock Fiber Spectra -- broadband colors and spectral mocks including dust attenuation effects \citep{nelson18a}.
  \item SKIRT Synthetic Images and Optical Morphologies -- dust radiative-transfer calculations using SKIRT to obtain broadband images, automated morphological measurements \citep{rodriguezgomez18}.
  \item DisPerSE Cosmic Web -- topological classification of the volume into sheets, filaments, nodes, and voids (\textcolor{blue}{Duckworth et al. in prep}).
  \item Particle-level lightcones -- in a variety of configurations, from small field of view `deep fields' to all-sky projections, across the different matter components, to facilitate lensing, x-ray, Sunyaev-Zeldovich, and related explorations (\textcolor{blue}{Giocoli et al. in prep}).
\end{enumerate}

Several of these were previously available for the original Illustris simulation and will be re-computed for TNG. We would plan to provide a number of `pre-defined' galaxy samples, particularly with respect to common observational selection techniques, current and/or upcoming surveys, and other distinct classes of interest. This can include, for example, red versus blue galaxies, luminous red galaxies (LRGs) and emission-line galaxies (ELGs) of SDSS, damped Lyman-alpha (DLA) host halos, and ultra-diffuse or low surface brightness (LSB) galaxies. Such samples would facilitate rapid comparisons to certain types of observational samples, and can be included as supplementary data catalogs as they become available.


\section{Data Access} \label{sDataAccess}

There are three complementary ways to access and analyze TNG data products.

\vspace{0.5em}
\begin{enumerate}
  \item \textbf{(Local data, local analysis).} Raw files can be directly downloaded, and example scripts are provided as a starting point for local analysis.
  \item \textbf{(Remote data, local analysis).} The web-based API can be used, either through a web browser or programmatically in a script, to perform search, data extraction, and visualization tasks, followed by local analysis of these derivative products.
  \item \textbf{(Remote data, remote analysis).} A web-based JupyterLab (or Jupyter notebook) session can be instantiated to explore the data, develop analysis scripts with persistent storage, run data-intensive and compute-intensive tasks, and make final plots for publication.
\end{enumerate}
\vspace{0.5em}

These different approaches can be combined. For example, by downloading the full redshift zero group catalog to perform a complex search which cannot be easily done with the API. After determining a sample of interesting galaxies (i.e. a set of subhalo IDs), one can then extract their individual merger trees (and/or raw particle data) without needing to download the full simulation merger tree (or a full snapshot).

These approaches are described below, while ``getting started'' tutorials for several languages (currently: Python, IDL, and Matlab) can be found online.

\subsection{Direct File Download and Example Scripts}

\textbf{Local data, local analysis.} All of the primary outputs of the TNG simulations are released in HDF5 format, which we use universally for all data products. This is a portable, self-describing, binary specification (similar to FITS), suitable for large numerical datasets. File access libraries, routines, and examples are available in all common computing languages. We typically use only basic features of the format: attributes for metadata, groups for organization, and large datasets containing one and two dimensional numeric arrays. To maintain reasonable filesizes for transfer, most outputs are split across multiple files called ``chunks''. For example, each snapshot of TNG100-1 is split into 448 sequentially numbered chunks. Links to the individual file chunks for a given simulation snapshot or group catalog are available under their respective pages on the main data release page.

The provided example scripts (in IDL, Python, and Matlab) give basic I/O functionality, and we expect they will serve as a useful starting point for writing any analysis task, and intend them as a `minimal working examples' which are short and simple enough that they can be quickly understood and extended. For a getting-started guide and reference, see the online documentation.

\subsection{Web-based API}

\textbf{Remote data, local analysis.} For TNG we enhance the web-based interface (API) introduced with the original Illustris simulation, augmented by a number of new features and more sophisticated functionality. At its core, the API can respond to a variety of user requests and queries. It provides a well-defined interface between the user and simulation data, and the tools it provides are independent, as much as possible, from any underlying details of data structure, heterogeneity, storage format, and so on. The API can be used as an alternative to downloading large data files for local analysis. Fundamentally, the API allows a user to \textbf{search}, \textbf{extract}, \textbf{visualize}, or \textbf{analyze} a simulation, a snapshot, a group catalog, or a particular galaxy/halo. By way of example, the following requests can be handled by the current API:

\vspace{0.5em}
\begin{itemize}[leftmargin=2em]
\setlength\itemsep{0em}
  \item Search across subhalos with numeric range(s) over any field(s) present in the Subfind catalogs.
  \item Retrieve a snapshot cutout for all the particles/cells within a given subhalo/halo, optionally restricted to a subset of specified particle/cell type(s) and fields(s).
  \item Retrieve the complete merger history or main branches for a given subhalo.
  \item Download subsets of snapshot files, containing only specified particle/cell type(s), and/or specific field(s) for each type.
  \item Traverse links between halos and subhalos, for instance from a satellite galaxy, to its parent FoF halo, to the primary (central) subhalo of that group, as well as merger tree progenitor/descendant connections.
  \item Render visualizations of any field(s) of different components (e.g. dark matter, gas, stars) of a particular halo/subhalo.\footnote{\label{newfeat}New feature in the TNG data release.}
  \item Download actual data from such a halo/subhalo visualization, e.g. maps of projected gas density, O VI column density, or stellar luminosity in a given band.\textsuperscript{\textcolor{blue}{[\ref{newfeat}]}}
  \item Render a static visualization of the complete merger tree (assembly history) of any subhalo.\textsuperscript{\textcolor{blue}{[\ref{newfeat}]}}
  \item Plot the relationship between quantities in the group catalogs, e.g. fundamental scaling relations such as the star-forming main sequence of TNG.\textsuperscript{\textcolor{blue}{[\ref{newfeat}]}}
  \item Plot tertiary relationships between group catalog quantities, e.g. the dependence of gas fraction on offset from the main sequence.\textsuperscript{\textcolor{blue}{[\ref{newfeat}]}}
\end{itemize}
\vspace{0.5em}

The IllustrisTNG data access API is available at the following permanent URL:

\vspace{0.5em}
\begin{itemize}[leftmargin=2em]
\item[] \url{http://www.tng-project.org/api/}
\end{itemize}
\vspace{0.5em}

\noindent For a getting-started guide for the API, as well as a cookbook of common examples and the complete reference, see the online documentation.

\subsection{Remote Data Analysis}

\textbf{Remote data, remote analysis.} Coincident with the TNG public data release we introduce a new, third option for working with and analyzing large simulation datasets. Namely, an online, browser-based scientific computing environment which enables researchers' computations to ``be brought to'' the data. It is similar in spirit to the NOAO Data Lab \citep{fitzpatrick14} and SciServer services \citep{raddick17}, i.e. simultaneously hosting petabyte-scale datasets as well as a full-featured analysis platform and toolset. This alleviates the need to download any data, or run any calculations locally, thereby facilitating broad, universal, open access to large cosmological simulation datasets such as TNG.

To enable this functionality we make use of extensive development on Jupyter and JupyterLab over the last few years. JupyterLab is the evolution of the Jupyter Notebook \citep{kluyver16}, previously called IPython \citep{perez07}. It is a next-generation, web-based user interface suitable for scientific data analysis. In addition to the previous `notebook' format, JupyterLab also enables a traditional workflow based around a collection of scripts on a filesystem, text editors, a console, and command-line execution. It provides an experience nearly indistinguishable from working directly on a remote computing cluster via SSH.

Computation is language agnostic, as `kernels' are supported in all common languages, including Python 2.x, Python 3.x, IDL, Matlab, R, and Julia. Development, visualization, and analysis in any language or environment practically available within a Linux environment is possible, although we focus at present on Python 3.x support.

Practically, this service enables direct access to one of the complete mirrors of the Illustris[TNG] data, which is hosted at the Max Planck Computing and Data Facility (MPCDF) in Germany. Users can request a new, on-demand JupyterLab instance, which is launched on a system at MPCDF and connected to the user web browser. All Illustris[TNG] data is then directly available for analysis. A small amount of persistent user storage is provided, so that under-development scripts, intermediate analysis outputs, and in-progress figures for publication all persist across sessions. Users can log out and pick up later where they left off. A base computing environment is provided, which can be customized as needed (e.g. by installing new python packages with either \texttt{pip} or \texttt{conda}). Users can synchronize their pre-existing tools, such as analysis scripts, with standard approaches (\texttt{git}, \texttt{hg}, \texttt{rsync}) or via the JupyterLab interface. Results, such as figures or data files, can be viewed in the browser or downloaded back to the client machine with the same tools.

For security and resource allocation, users must specifically request access to the JupyterLab TNG service. At present we anticipate providing this service on an experimental (beta) basis, and only to active academic users.

\subsection{Further Online Tools}

\subsubsection{Subhalo Search Form}

We provide the same, simple search form to query the subhalo database as was available in the Illustris data release. It exposes the search capabilities of the API in a user-friendly interface, enabling quick exploration without the need to write a script or URL by hand. As examples, objects can be selected based on total mass, stellar mass, star formation rate, or gas metallicity. The tabular output lists the subhalos which match the search, along with their properties. In addition, each result contains links to a common set of API endpoints and web-based tools for inspection and visualization.

\subsubsection{Explore: 2D and 3D}

The 2D Explorer and 3D Explorer interfaces are experiments in the \textit{interactive} visualization and exploration of large data sets such as those generated by the IllustrisTNG simulations. They both leverage the approach of thin-client interaction with derived data products. The 2D Explorer exposes a Google Maps$-$like tile viewer of pre-computed imagery from a slice of the TNG300-1 simulation at redshift zero, similar to the original Illustris explorer. Multiple views of different particle types (gas, stars, dark matter, and blackholes) can be toggled and overlaid, which is particularly useful in exploring the spatial relationships between different phenomena of these four matter components.

\begin{figure*}[htb!]
\centering
\includegraphics[angle=0,width=3.2in]{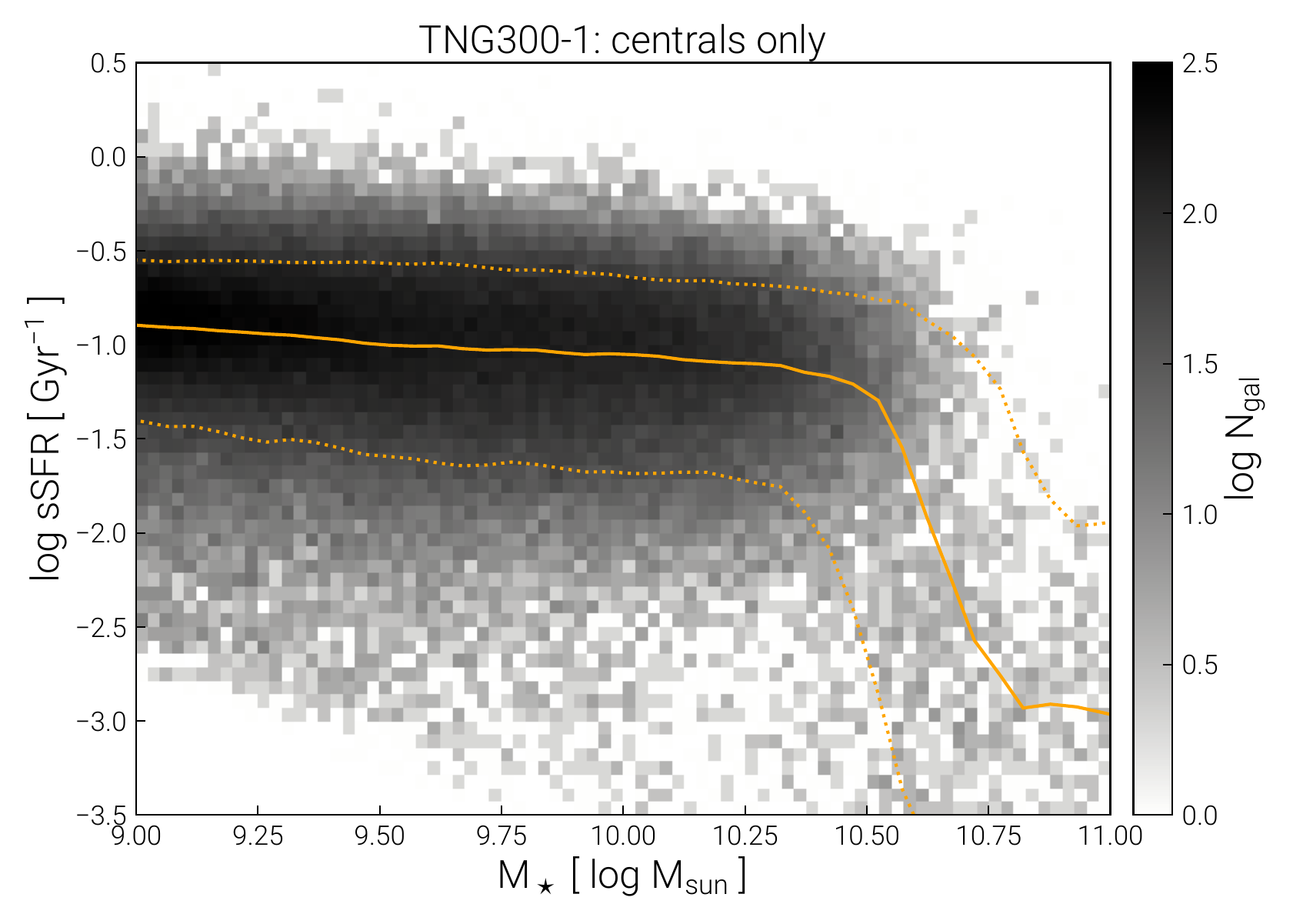}
\includegraphics[angle=0,width=3.2in]{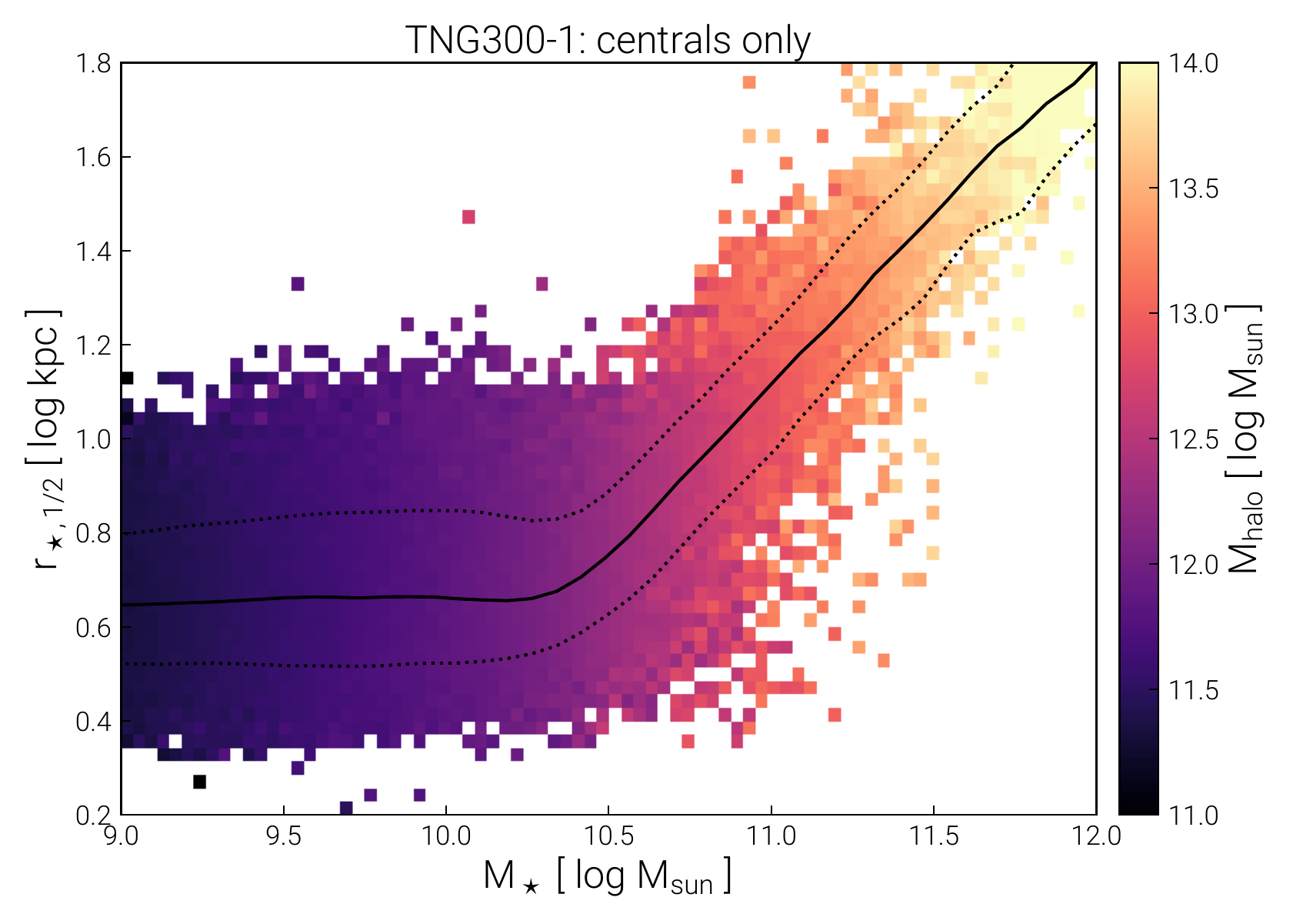}
\includegraphics[angle=0,width=3.2in]{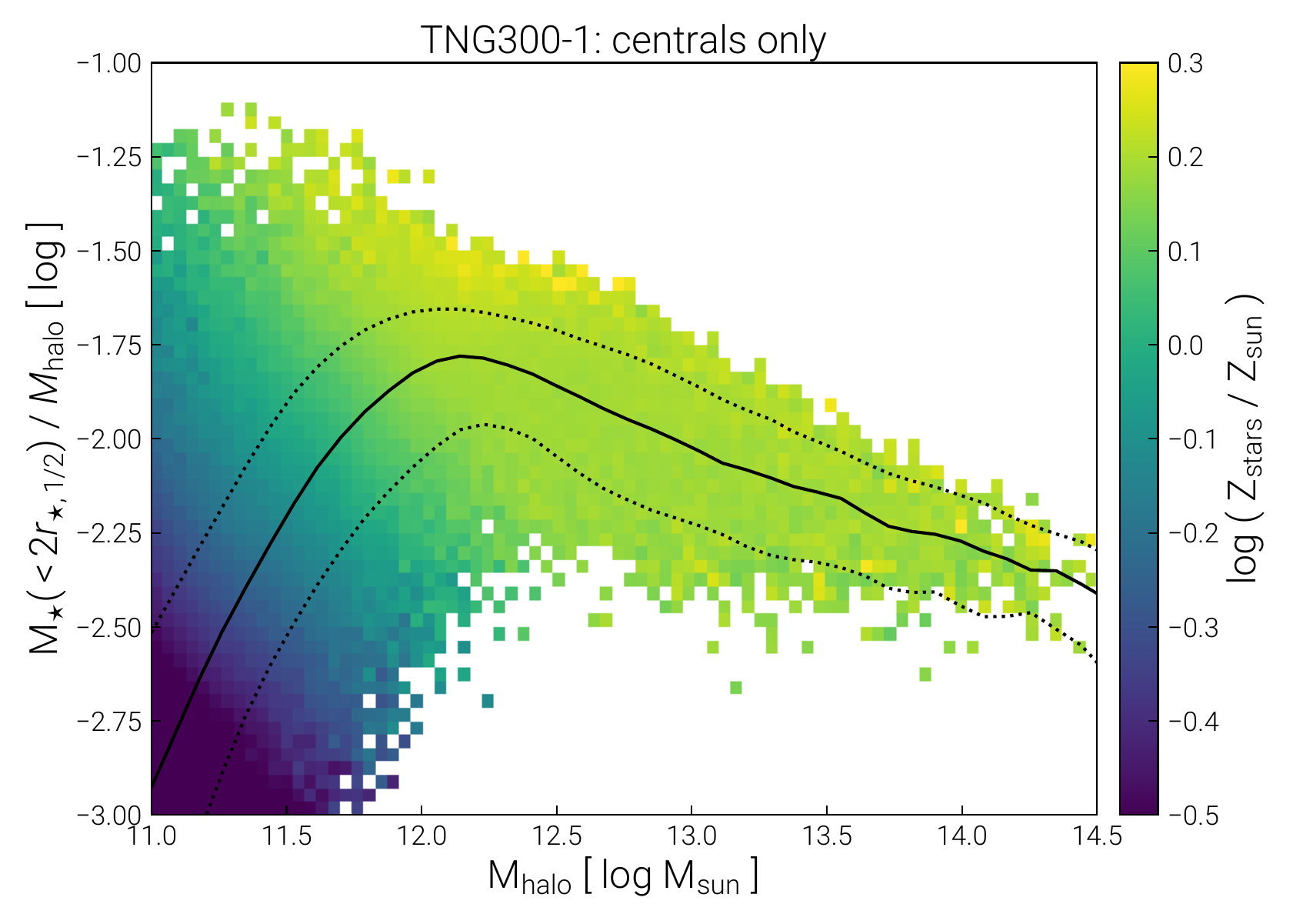}
\includegraphics[angle=0,width=3.2in]{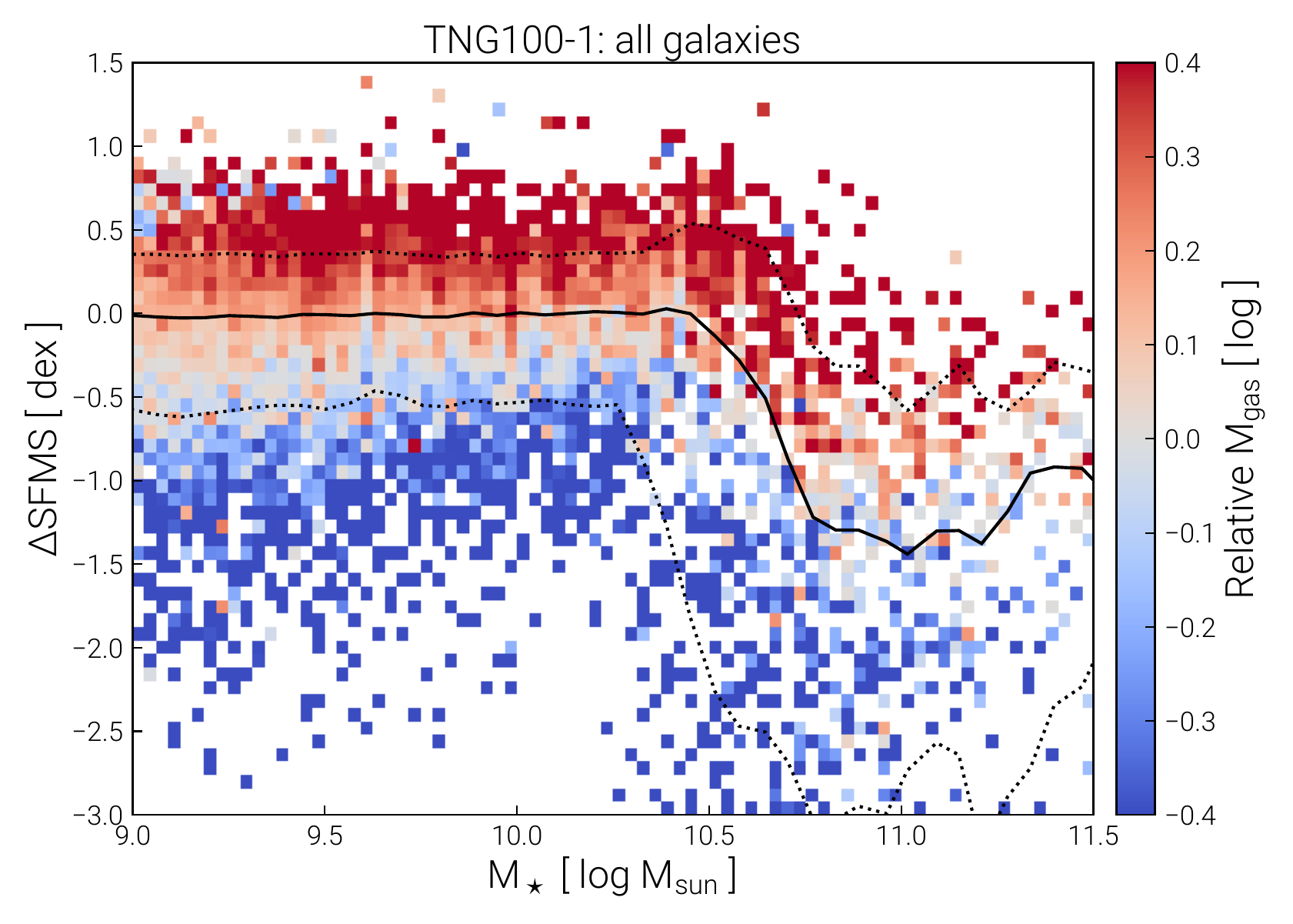}
\caption{ Four examples of exploratory plots for common scaling relations, galaxy trends, and other relationships between properties of the objects in the group catalogs, galaxies and halos, for TNG300-1 and TNG100-1 at $z=0$. Made using the web-based API functionality.
 \label{fig_plotex}} 
\end{figure*}

The 3D Explorer introduces a new interface, showing a highly derivative (although spatially complete) view of an entire snapshot. That is, instead of particle-level information, we facilitate interactive exploration of the group catalog output in three-dimensional space. This allows users to rotate, zoom, and move around the cubic box representing the simulation domain, where the largest dark matter halos are represented by wireframe spheres of size equal to their virial radii, while the remaining smaller halos are represented by points. User selection of a particular halo, via on-click ray cast and sphere intersection testing, launches an API query and returns the relevant halo information and further introspection links. At present, both Explorers remain largely proof of concept interfaces for how tighter integration of numeric, tabular, and visual data analysis components may be combined in the future for the effective exploration and analysis of large cosmological datasets \citep[see also][and the Dark Sky simulation]{dykes18}.

\subsubsection{Merger Tree Visualization}

In the Illustris data release we demonstrated a rich-client application built on top of the API, in the form of an interactive visualization of merger trees. The tree is vector based, and client side, so each node can be interacted with individually. The informational popup provides a link, back into the API, where the details of the selected progenitor subhalo can be interrogated. This functionality is likewise available for all new simulations. Furthermore, we have added a new, static visualization of the complete merger tree of a subhalo. This allows a quick overview of the assembly history of a given object, particularly its past merger events and its path towards quiescence. In the fiducial configuration, node size in the tree is scaled with the logarithm of total halo mass, while color is mapped to instantaneous sSFR.

\subsubsection{Plot Group Catalog}

The first significant new feature of the API for the TNG public data release is a plotting routine to examine the group catalogs. Since the objects in the catalogs are either galaxies or dark matter halos, plotting the relationships among their various quantities is one of most fundamental explorations of cosmological simulations. Classically observed scaling relations, such as Tully-Fisher (rotation velocity vs. stellar mass), Faber-Jackson (stellar velocity dispersion vs. luminosity), the stellar size-mass relation, the star-formation main sequence, or the Magorrian relation (blackhole mass versus bulge mass) are all available herein. Such relations can be used to assess the outcome of the simulations by comparison to observational data. More complex relations, those involving currently unobserved properties of galaxies/halos, and/or those only currently observed with very limited statistics or over limited dynamic range, represent a powerful discovery space and predictive regime for simulations such as TNG. At the level of the galaxy (or halo) population, i.e. with tens to hundreds of thousands of simulated objects, many such relationships reveal details of the process of galaxy formation and evolution, as well as the working mechanisms of the physical/numerical models.

\begin{figure*}[htb!]
\centering
\includegraphics[angle=0,width=1.59in]{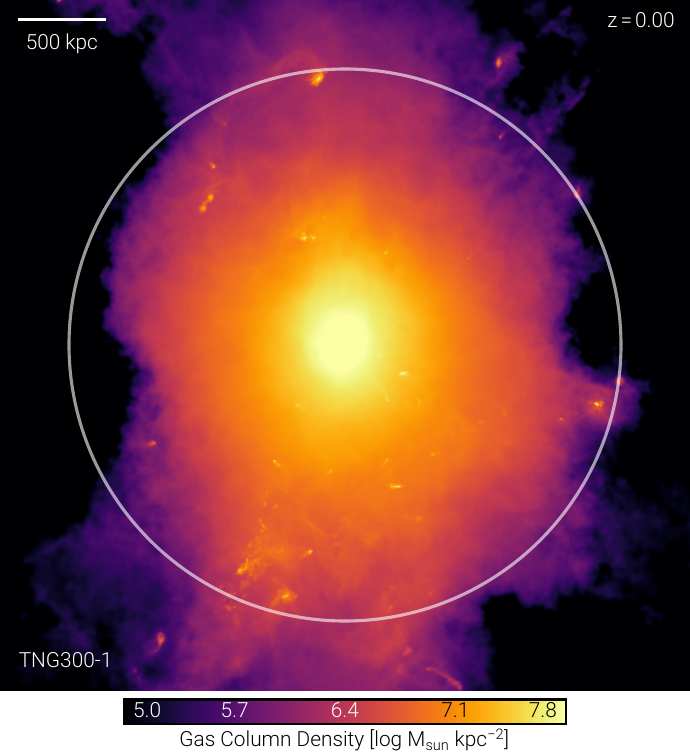}
\includegraphics[angle=0,width=1.59in]{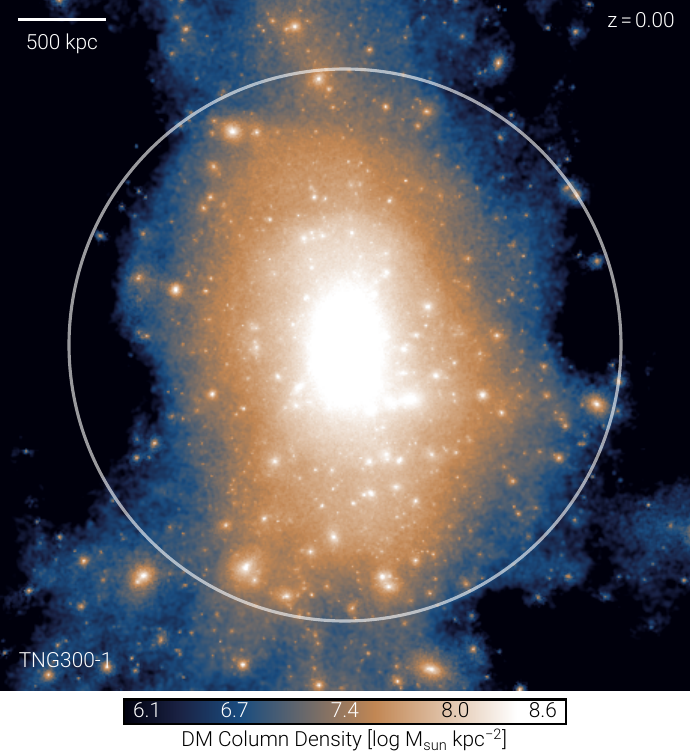}
\includegraphics[angle=0,width=1.59in]{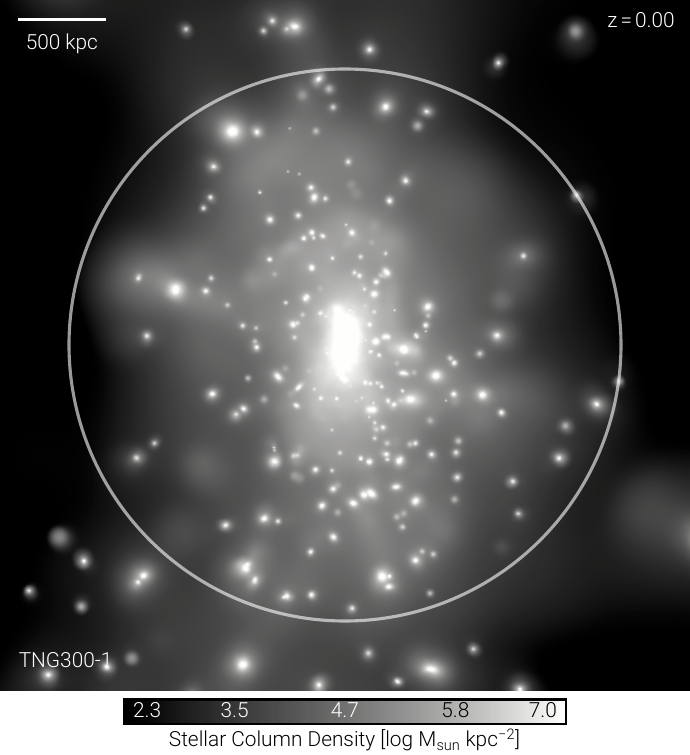}
\includegraphics[angle=0,width=1.59in]{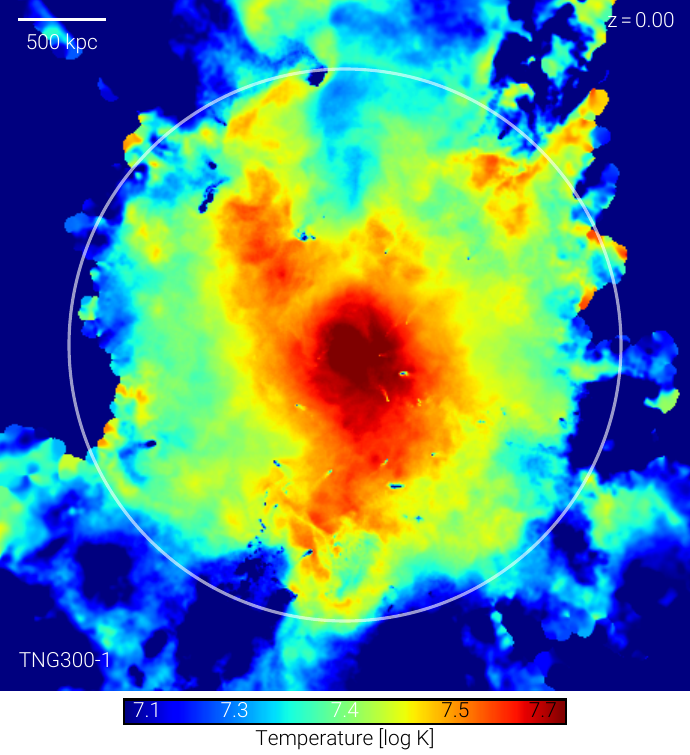}
\includegraphics[angle=0,width=1.59in]{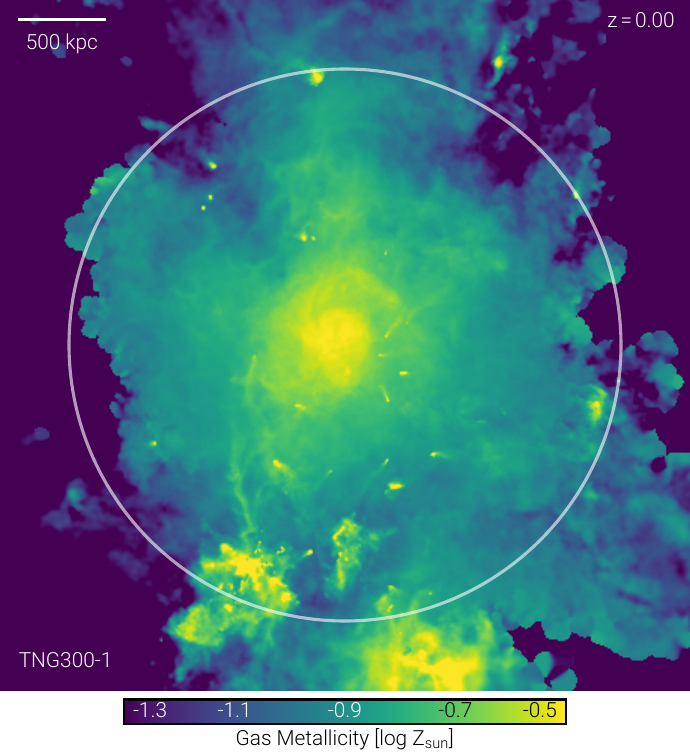}
\includegraphics[angle=0,width=1.59in]{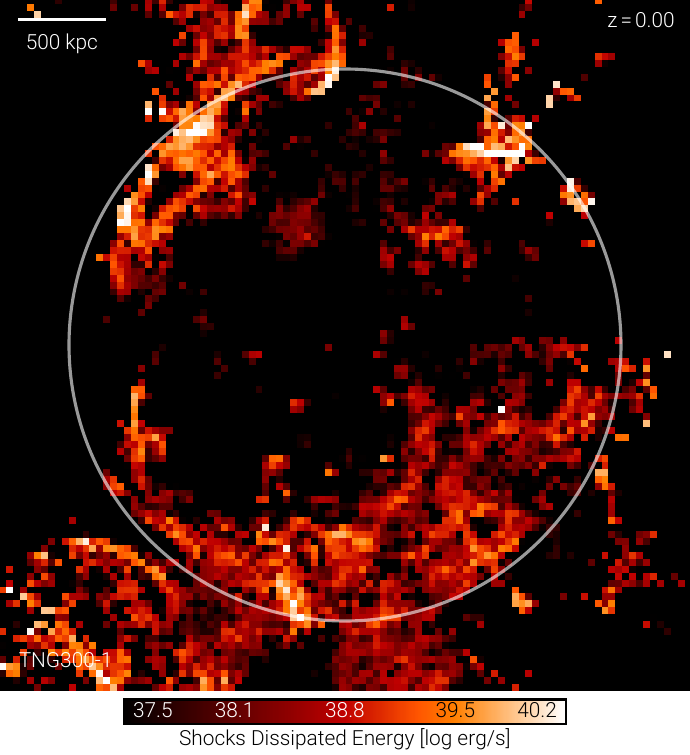}
\includegraphics[angle=0,width=1.59in]{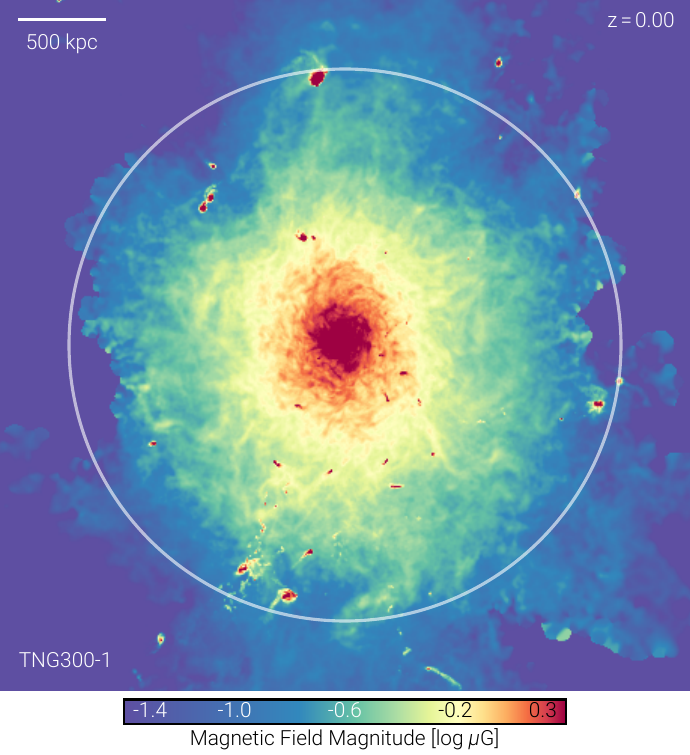}
\includegraphics[angle=0,width=1.59in]{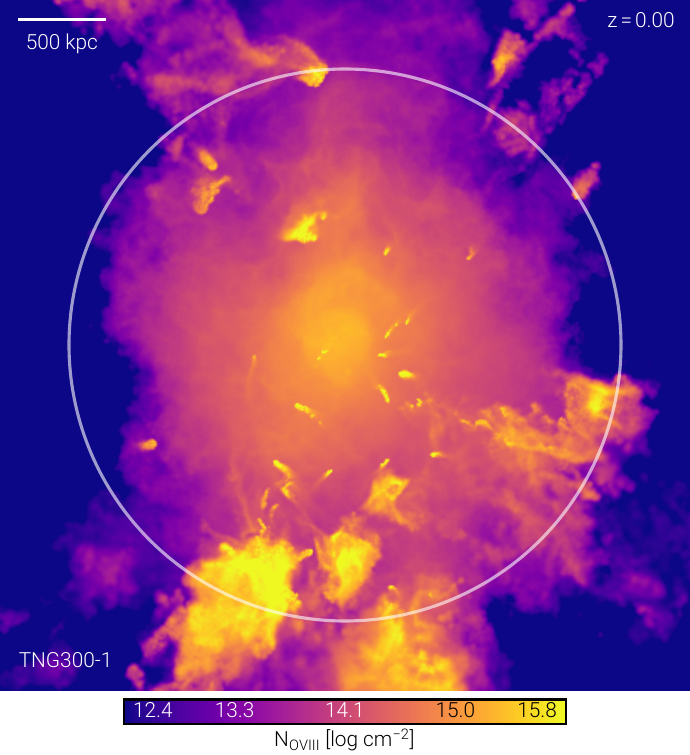}

\includegraphics[angle=0,width=1.59in]{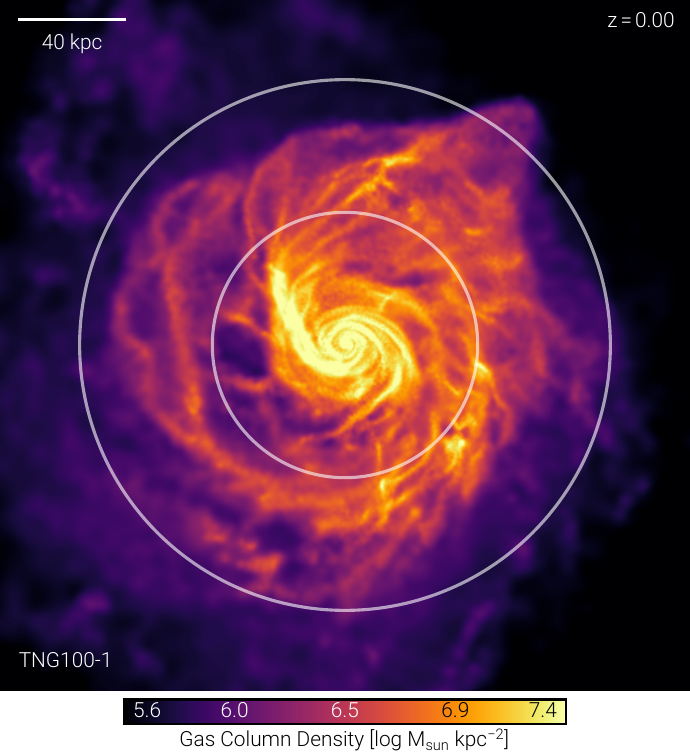}
\includegraphics[angle=0,width=1.59in]{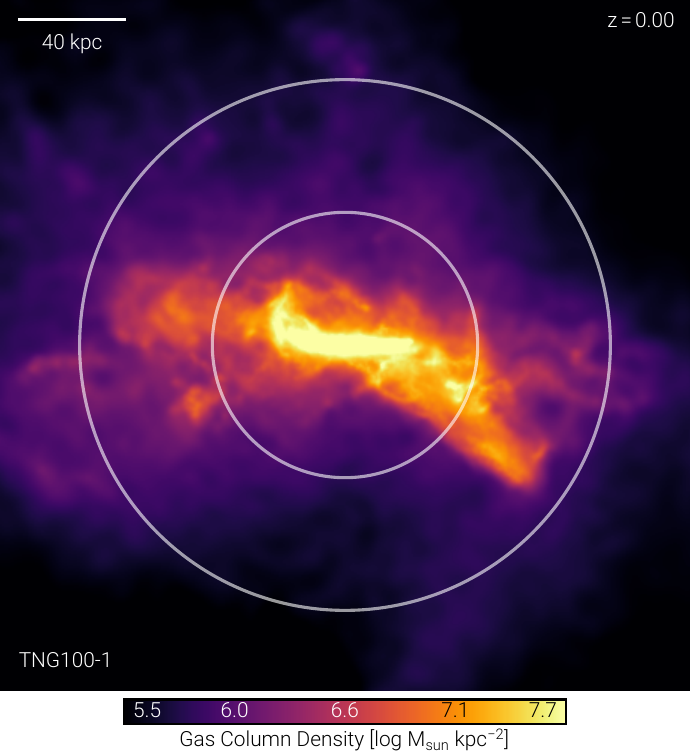}
\includegraphics[angle=0,width=1.59in]{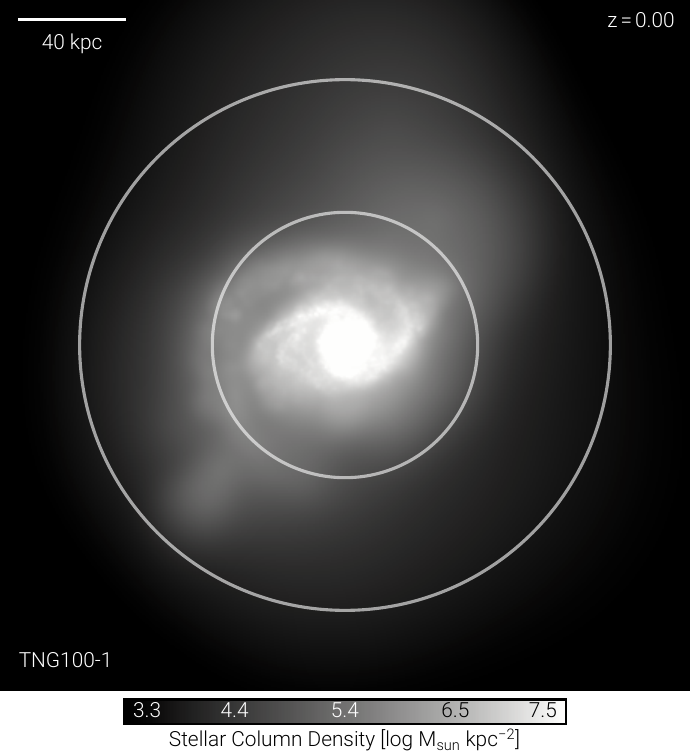}
\includegraphics[angle=0,width=1.59in]{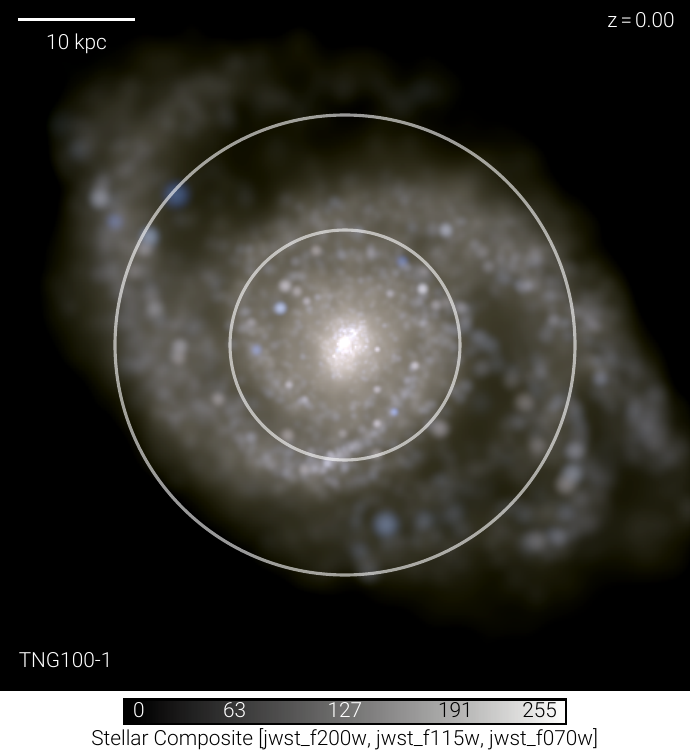}
\includegraphics[angle=0,width=1.59in]{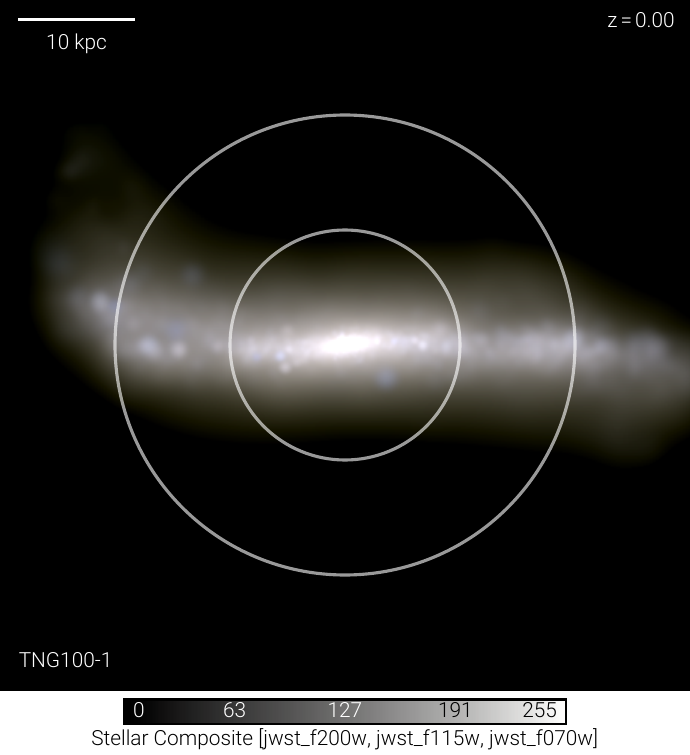}
\includegraphics[angle=0,width=1.59in]{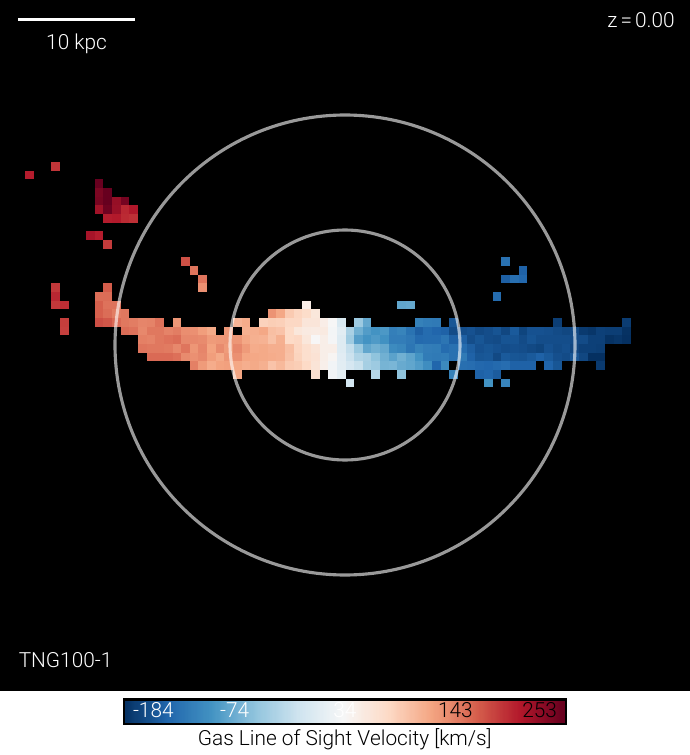}
\includegraphics[angle=0,width=1.59in]{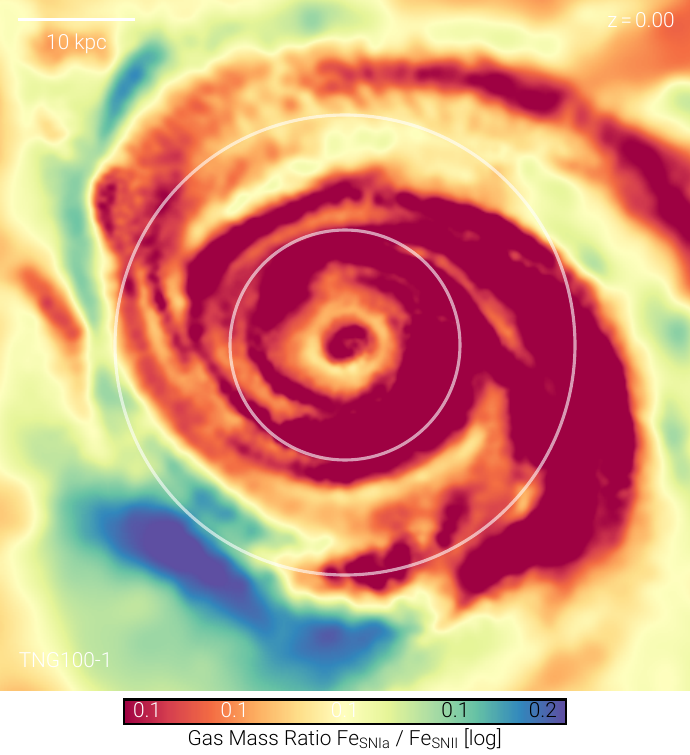}
\includegraphics[angle=0,width=1.59in]{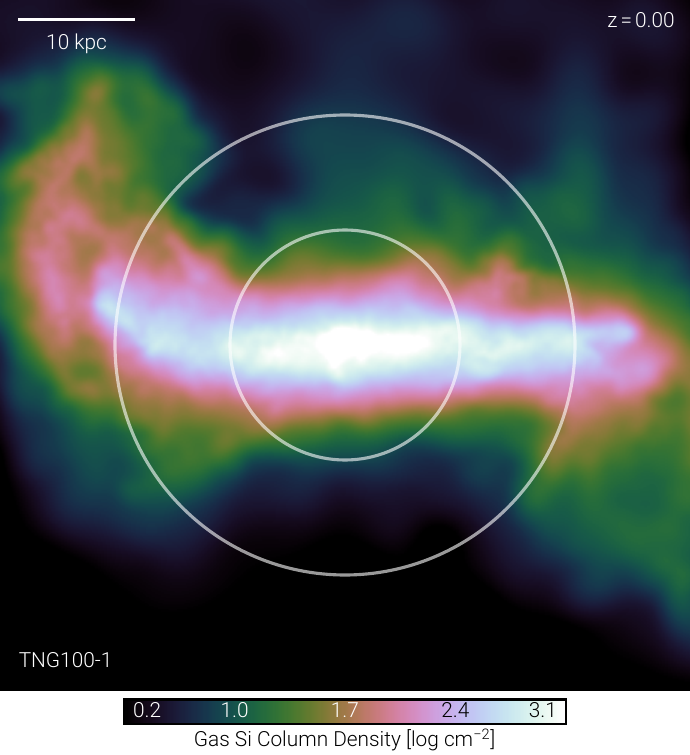}
\caption{ Example of halo-scale and galaxy-scale visualizations from TNG300-1 and TNG100-1, made using the web-based API functionality, viewing the dark matter, gas, and stars. The top eight panels show the 20th most massive halo of TNG300-1 at $z=0$ (circle indicating $r_{\rm vir}$). The bottom eight panels show face-on and edge-on views of subhalo 468590 of TNG100-1 at $z=0$ (circles indicating $r_{\rm 1/2,\star}$ and $2r_{\rm 1/2,\star}$).
 \label{fig_visex}} 
\end{figure*}

The `group catalog plotter' is an API endpoint which returns publication quality figures (e.g. PNG or PDF outputs). In Figure \ref{fig_plotex} we show several examples of its output, taken from TNG300-1 and TNG100-1 at $z=0$. Many options exist to control the behavior and structure of the plots, all of which are detailed in the online documentation. As for the subhalo search form, we also provide a new web-based interface to assist in interactively constructing plots from this service. Fundamentally, the quantities to be plotted against each other on the x- and y-axes can be selected. In this case, a two-dimensional histogram showing the density of subhalos in this space is overlaid with the median relation and bounding percentiles. Optionally, a third quantity can be added, which is then used to color every pixel in the histogram according to a user-defined statistic (e.g. median) of all the objects in that bin. For example, plotting the stellar-mass halo-mass relation, colored by stellar metallicity, reveals one reason for the scatter in this relation. This third quantity can optionally be normalized relative to the median value at each x-axis value (e.g. as a function of stellar mass), highlighting the `relative' deviation of that property compared to its evolving median value. The types of subhalos included can be chosen, for example selecting only centrals or only satellite galaxies, and the subhalos to be included can be filtered based on numeric range selections on a fourth quantity. We expect that this tool will enable rapid, initial exploration of interesting relationships among galaxy (or halo) integral properties, and serve as a starting point for more in-depth analysis \citep[see also][]{desouza15}. Complete usage documentation is available online.

\subsubsection{Visualize Galaxies and Halos}

The second significant new feature of the API for the TNG public data release is an on-demand visualization service. Primarily, this API endpoint renders projections of particle-level quantities (of gas cells, dark matter particles, or stellar particles) for a given subhalo or halo. For example, it can produce gas column density projections, gas temperature projections, stellar line-of-sight velocity maps, or dark matter velocity dispersion maps. Its main rendering algorithm is based on the standard SPH kernel projection technique, with adaptive kernel sizes for all particle types, although alternatives are available. In Figure \ref{fig_visex} we show several examples of output, at both the halo-scale (circle indicating virial radius), and the galaxy scale (outer circle showing twice the stellar half mass radius).

The visualization service returns publication quality figures (e.g. PNG or PDF outputs). It can also return the raw data used to construct any image, in scientifically accurate units (HDF5 output). For instance, a user can request not only an image of the gas density projection of an ongoing galaxy merger, but also the actual grid of density values in units of e.g. $\rm{M}_{\odot}\, \rm{kpc}^{-2}$. Numerous options exist to control the behavior of the rendered projections, as well as the output style, all of which are detailed in the online documentation. All parameters of the rendering can be specified -- as an example, the view direction can be a rotation into face-on or edge-on orientations. Most properties available in the snapshots can be visualized, for any galaxy/halo, at any snapshot, for any run.

Beyond snapshot level information, the visualization service currently has two more advanced features. First, it is coupled to the {\sc CLOUDY} photoionization code \citep{ferland17}, following \cite{nelson18b}. This enables ionic abundance calculations for gas cells on the fly. For example, a user can request a column density map of the O VI or C IV ions. All relevant atoms are supported, assuming solar abundances for non-tracked elements, typically up to the tenth ionization state (Al, Ar, Be, B, Ca, C, Cl, Cr, Co, Cu, F, He, H, Fe, Li, Mg, Mn, Ne, Ni, N, O, P, K, Sc, Si, Na, S, Ti, V, Zn). Emission line luminosities are also available -- a surface brightness map of metal-line emission from O VIII at 22.1012\AA, for example. Secondly, this service is also coupled to the {\sc FSPS} stellar population synthesis code \citep{conroy09,conroy10} through {\sc python-fsps} \citep{foremanmackey14}, following \cite{nelson18a}. This enables emergent stellar light calculations for stellar population particles on the fly, with optional treatments of dust attenuation effects. For example, a user can request a map of stellar surface brightness, or luminosity, either rest frame or observed frame, for any of the $\sim$140 available bands, including common filters on surveys/instruments such as SDSS, DES, HST, and JWST.

We expect that this tool will enable rapid, initial exploration of many interesting facets of galaxies and halos across the simulations, and serve as a starting point for more in-depth analysis. We caution that, used improperly, this tool can easily return nonsensical results (e.g. requesting OI emission properties from ISM gas), and users should understand the relevant scientific limitations. In this particular case, we refer to the effective two-phase ISM model used in TNG \citep{spr03} which intentionally avoids resolving the cold, dense phases of the ISM. Complete usage documentation is available online.


\section{Scientific Remarks and Cautions} \label{sRemarks}

In the original Illustris simulation we identified a number of non-trivial issues in the simulated galaxy and halo populations in comparison to observational constraints \citep[see][for a summary]{nelson15b}. These disagreements motivated a series of important caveats against drawing certain strong scientific conclusions in a number of regimes.

In contrast, our initial explorations of TNG (specifically, of the TNG100-1 and TNG300-1 simulations) have revealed no comparably significant tensions with respect to observable comparisons. With this data release we invite further detailed observational comparisons and scrutiny. The TNG simulations have been shown to realistically resolve numerous aspects of galactic structure and evolution, including many internal properties of galaxies (though, clearly, not all) as well as their co-evolution within the cosmic web of large-scale structure (see Section \ref{sec_earlyresults}). TNG reproduces various observational details and scaling relations of the demographics and properties of the galaxy population, not only at the present epoch ($z=0$), but also at earlier times (see likewise Section \ref{sec_earlyresults}). This has been achieved with a physically plausible although necessarily simplified galaxy formation model. The TNG model is intended to account for most, if not all, of the primary processes that are believed to be important for the formation and evolution of galaxies.

\subsection{IllustrisTNG: Possible Observational Tensions}

We therefore do not specifically caution against the use of TNG in any of the regimes where the original Illustris simulation was found to be less robust. However, the enormous spatial and temporal dynamic range of these simulations, as well as the multi-scale, multi-physics nature of the complex physical phenomena involved, implies modeling approximations and uncertainties. Early comparisons of TNG against observations have identified a number of interesting regimes in which possible tensions exist. 

Our ability to make any stronger statement is frequently limited by the complexity of the observational comparison, i.e. the need to accurately reproduce (or `mock') the observational measurement closely and with care. In the qualitative sense, however, these regimes may plausibly indicate areas where the TNG model has shortcomings or is less physically realistic. It will be helpful for any user of the public data to be aware of these points, which should be carefully considered when advancing strong scientific conclusions or making claims based on observational comparisons.
Possible tensions of interest include the following:

\vspace{0.5em}
\begin{enumerate}[label=(\Roman*)]
\setlength\itemsep{0em}
  \item The simulated stars in Milky Way-like galaxies are too alpha-enhanced in comparison to observations of the Milky Way \citep{naiman18}.
  
  \item The Eddington ratio distributions of massive blackholes ($> 10^9\, {\rm M}_{\odot}$) at $z=0$ are dominated in TNG by low accretion rates, generically far below the Eddington limit; recent observations favor at least some fraction of higher accretion rate massive blackholes. This is reflected in a steeper hard X-ray AGN luminosity function at $1 \lesssim z \lesssim 4$ \citep{habouzit18}.
  
  \item TNG galaxies may have a weaker connection between galaxy morphology and color than observed at $z\sim0$, reflected in a possible excess of red disk-like galaxies in the simulations \citep{rodriguezgomez18}, although see \textcolor{blue}{Tachella et al. (in prep)}.
  
  \item TNG galaxies exhibit a somewhat sharper trend than observations in quenched fraction vs. galaxy stellar mass for $M_\star \in 10^{10-11}{\rm M}_{\odot}$ \citep{donnari19}, and similarly in the relation between sSFR and $M_{\rm BH}$ at low redshift (\textcolor{blue}{Terrazas et al. in prep}).
  
  \item The locus of the galaxy star-forming main sequence is below the face-value observed SFMS at $1 \lesssim z \lesssim 2$, modulo known inconsistencies with e.g. the observed stellar mass function \citep{donnari19}.
  
  \item Similarly, the H$_2$ mass content of massive TNG galaxies at $z=1-3$ may be lower than implied by ALMA observations \citep{popping19} and sub-mm galaxy demographics (\textcolor{blue}{Hayward et al. in prep}).
  
  \item The DM fractions within massive elliptical galaxies at $z=0$ are consistent with observations at large galactocentric distances, but may be too high within their effective radii \citep{lovell18} and likewise are tentatively higher than values inferred from observations of massive $z=2$ star forming galaxies \citep[][and \textcolor{blue}{{\"U}bler et al. in prep}]{lovell18}.
\end{enumerate}
\vspace{0.5em}

With respect to points (III)$-$(IV) there is, in general, an interesting transitional mass regime (galaxy stellar mass $\sim 10^{10.5} {\rm M}_{\odot}$) where central blue vs. red galaxies or star-forming vs. quiescent galaxies co-exist: this reflects the effective quenching mechanism of the TNG model based on SMBH feedback \citep{nelson18a, weinberger18} but how precisely such transitional galaxies differ also in other structural and kinematical properties still requires careful examination and consideration.

Note that for the items in this list we have not included more specific quantification of observed tension (i.e. $\chi^2$ or fractional deviation values) -- the referenced papers contain more discussion. On the one side, not all observational results are in agreement among each other, making quantitative statements necessarily partial; nor observational statements of different galaxy properties are necessarily consistent within one another, especially across cosmic times. On the other side, excruciating care is often necessary to properly map simulated variables into observationally-derived quantities.

\subsection{Numerical Considerations and Issues}

To better inform which features of the simulations are robust when making science conclusions, we note the following points related to numerical considerations:

\textbf{1. SubhaloFlag.} Not all objects in the Subfind group catalogs should be considered `galaxies'. In particular, not all satellite subhalos have a cosmological origin, in the sense that they may not have formed due to the process of structure formation and collapse. Rather, some satellite subhalos will represent fragments or clumps produced through baryonic processes (e.g. disk instabilities) in already formed galaxies, and the Subfind algorithm cannot \textit{a priori} differentiate between these two cases. Such non-cosmological objects are typically low mass, small in size, and baryon dominated (i.e. with little or no dark matter), residing at small galactocentric distances from their host halos, preferentially at late times ($z<1$). These objects may appear as outliers in scatter plots of typical galaxy scaling relations, and should be considered with care.

We have added a {\sc SubhaloFlag} field to the group catalogs to assist in their identification, which was constructed as follows. First, a variant of the SubLink merger tree was used which tracks baryonic, rather than dark matter, particles -- namely, star-forming gas cells and stars \citep{rodriguezgomez15}. The algorithm is otherwise the same, with the same weighting scheme for determining descendants/progenitors, except that this ``SubLinkGal'' tree allows us to track subhalos which contain little or no dark matter. 

Then, we flag a subhalo as non-cosmological if all the following criteria are true: (i) the subhalo is a satellite at its time of formation, (ii) it forms within 1.0 virial radii of its parent halo, and (iii) its dark matter fraction, defined as the ratio of dark matter mass to total subhalo mass, at the time of formation of the subhalo, is less than 0.8.

These are relatively conservative choices, implying a low false positive rate. On the other hand, some spurious subhalos may not be flagged under this definition. A much more aggressive criterion would be to flag a subhalo if its instantaneous dark matter fraction is low, e.g. less than 10\% \citep[as used in e.g.][]{genel18,pillepich18a}. Such a selection will result in a purer sample, with less contaminating subhalos, but will also exclude more genuine galaxies, such as those which have undergone extensive (i.e. physical) stripping of their dark matter component during infall. We encourage users to enforce the provided SubhaloFlag values as a default, but to carefully consider the implications and details, particularly for analyses focused on satellite galaxy populations or dark-matter deficient systems.

\textbf{2. Gas InternalEnergy Corrections.} In all TNG simulations, the time-variable UV-background radiation field \citep[][FG11 version]{fg09} is enabled only for $z < 6$. Therefore, the ionization state of the IGM above redshift six should be studied with caution, as the usual density-temperature relation will not be present. Two further technical issues exist for the original {\sc InternalEnergy} field (i.e. gas temperature) of all TNG simulations. These have been corrected in post-processing, as described below, and the fiducial {\sc InternalEnergy} field of all snapshots in all TNG simulations has been rewritten with updated values. The original dataset has been renamed to {\sc InternalEnergyOld} for reference, although we do not recommend its use for any purpose.

The first issue is seen in the low-density, low-temperature regime of the intergalactic medium (IGM). Here, due to a numerical issue in the TNG codebase related to the Hubble flow across gas cells, spurious energy injection could occur in underdense gas. In practice, this only affects extremely low density IGM gas in equilibrium between adiabatic cooling and photoheating from the background. The result is a slight upwards curvature in the usual $(\rho,T)$ phase diagram. To correct this issue, we have used one of the TNG model variant boxes (with side length $25 \rm{Mpc}/h$ and $512^3$ resolution) which includes the fix for this issue. The adiabat was then identified in all TNG runs as well as in the corrected simulation by binning the density-temperature phase diagram and locating the temperature of peak gas mass occupancy as a function of density. A multiplicative correction $f_{\rm corr}$, taken as the ratio between the corrected and uncorrected linear gas temperatures, is then defined and applied as a function of density, for gas with physical hydrogen number density $< 10^{-6} a^{-3} \,\rm{cm}^{-3}$. We further restrict the correction to the low-temperature IGM by smoothly damping $f_{\rm corr}$ to unity by $10^{5.0}$ K as $\log T_{\rm corr} = \log T_{\rm orig} + \log(f_{\rm corr}) w(T_{\rm orig})$ with the window function $w(T_{\rm orig}) = 1 - [\rm{tanh}(5(T_{\rm orig}-5.0))+1]/2$. This issue manifests only towards low redshift, and for simplicity and clarity we apply this correction only for $z \leq 5$ (snapshots 17 and later).

The second issue arises for a very small fraction of low-temperature gas cells with $T < 10^4\,\rm{K}$, the putative cooling floor of the model. Here, due to a numerical issue in the TNG codebase related to the cosmological scaling of the energy source term in the Powell divergence cleaning of the MHD scheme \citep[right-most term in Eqn. 21 of][]{pakmor13}, spurious cooling could occur in gas with high bulk velocity and large, local divergence error ($|\nabla \vec{B}| > 0$). In practice, this affects a negligible number of cells which appear in the usual $(\rho,T)$ phase diagram with temperatures less than 10,000 K and densities between the star-formation threshold and four orders of magnitude lower. To correct this issue we simply update the gas temperature values, for all cells in this density range with $\log(T \rm{[K]}) < 3.9$, to the cooling floor value of $10^{4}\,\rm{K}$, near the background equilibrium value. As this issue also manifests only towards low redshift (being most problematic at intermediate redshifts $1 \lesssim z \lesssim 4$),  we likewise apply this correction only for $z \leq 5$ (snapshots 17 and later). 

Note that for both issues, we have verified in reruns of smaller volume simulations, by applying the fix in correspondingly corrected TNG model variant simulations, that no properties of galaxies or of the galaxy population are noticeably affected by these fixes.

\textbf{3. Unresolved ISM.} The multi-phase model of the interstellar medium in TNG (which is the same as in Illustris) is a necessarily coarse approximation of a complex physical regime. In particular, the cold neutral and molecular phases of the ISM are not resolved in the current generation of cosmological simulations like TNG; giant molecular clouds (GMCs) and the individual birth sites of massive star formation and, for example, the resultant nebular excitation is likewise not explicitly resolved. Modeling observables which arise in dense ISM phases (e.g. CO masses) should be undertaken with care.

The modeling of the star formation process is explicitly designed to reproduce the empirical Kennicutt-Schmidt relation, so the correlation between star formation rate and gas density, at the scale where this scaling is invoked, is not a predictive result. Star formation, as in all computational models of galaxy formation, proceeds at a numerical threshold density which is many orders of magnitude lower than the true density at which stars form. This threshold is $n_{\rm H} \simeq 0.1 \rm{cm}^{-3}$ in TNG, which may have consequences for the spatial clustering of young stars, as one example \citep{buck18}.

\textbf{4. Resolution Convergence.} Numerical convergence is a complex issue, and working with simulations at multiple resolutions can be challenging. Analysis which includes more than one TNG box at once (e.g. TNG100 and TNG300 together), or explicitly uses multiple realizations at different resolutions should carefully consider the issue of convergence. The degree to which various properties of galaxies or the simulation as a whole is converged depends on the specific property, as well as the mass regime, redshift, and so on. For example, see \cite{pillepich18a} for convergence of the stellar mass functions of TNG100 and TNG300, and details on a simple `resolution correction' procedure which may be desirable to apply, particularly when combining the results of multiple flagship boxes together into a single analysis.


\section{Community Considerations} \label{sCommunity}

\subsection{Citation Request}

To support proper attribution, recognize the effort of individuals involved, and monitor ongoing usage and impact, the following is requested. Any publication making use of data from the TNG100/TNG300 simulations should cite this release paper \citep{nelson19a} as well as the five works from the ``introductory paper series'' of TNG100/300, the order being random:

\vspace{0.5em}
\begin{itemize}
\setlength\itemsep{0.2em}
  \item \cite{pillepich18b} (stellar contents),
  \item \cite{springel18} (clustering),
  \item \cite{nelson18a} (colors),
  \item \cite{naiman18} (chemical enrichment),
  \item \cite{marinacci18} (magnetic fields).
\end{itemize}
\vspace{0.5em}

Any publication making use of the data from the TNG50 simulation should cite this release paper, as well as the two introductory papers of TNG50, the order being random: 

\vspace{0.5em}
\begin{itemize}
\setlength\itemsep{0.2em}
  \item \cite{nelson19b} (outflows),
  \item \cite{pillepich19} (structure \& kinematics).
\end{itemize}
\vspace{0.5em}

Finally, use of any of the supplementary data products should include the relevant citation. A full and up to date list is maintained on the TNG website. 

\subsection{Collaboration and Contributions}

The full snapshots of TNG50-1, TNG100-1, and especially those of TNG300-1, are sufficiently large that it may be prohibitive for most users to acquire or store a large number. We note that transferring $\sim\,$1.5 TB (the size of one full TNG100-1 snapshot) at a reasonably achievable 10 MB/s will take roughly 48 hours, increasing to roughly five days for a $\sim\,$4.1 TB full snapshot of TNG300-1. As a result, projects requiring access to full simulation datasets, or extensive post-processing computations beyond what are being made publicly available, may benefit from closer interaction with members of the TNG collaboration. 
      
We also welcome suggestions, comments, or contributions to the data release effort itself. These could take the form of analysis code, derived data catalogs, etc. For instance, interesting data products can be released as a ``supplementary data catalog''. Fast analysis routines which operate on individual halos/subhalos can be integrated into the API, such that the result can be requested on demand for any object.

\subsection{Future Data Releases}

We anticipate to release additional data in the future, for which further documentation will be provided online.

\subsubsection{Rockstar and Velociraptor}

We plan to derive and release different group catalogs, based on the {\sc Rockstar} \citep{behroozi13} and {\sc Velociraptor} \citep{elahi11} algorithms, and will provide further documentation at that time. Such group catalogs will identify different subhalo populations than found by the Subfind algorithm, particularly during mergers. The algorithm used to construct the `Consistent Trees' assembly histories also has fundamental differences to both {\sc LHaloTree} and {\sc SubLink}. This can provide a powerful comparison and consistency check for any scientific analysis. We also anticipate that some users will simply be more familiar with these outputs, or need them as inputs to other tools.

\subsubsection{Additional Simulations}

The flagship volumes of the IllustrisTNG -- TNG50, TNG100, and TNG300 -- are accompanied by an additional resource: a large number of `TNG Model Variation' simulations. Each modifies exactly one choice or parameter value of the base, fiducial TNG model. The variations cover every physical aspect of the model, including the stellar and blackhole feedback mechanisms, aspects of the star formation, as well as numerical parameters. They are invaluable in assessing the robustness of a physical conclusion to model changes, as well as in diagnosing the underlying cause or mechanism responsible for a given feature in the primary simulations. They were first presented in the \cite{pillepich18a} TNG methods paper, and used for example in \cite{nelson18b} to understand the improvement in OVI column densities, in \cite{lovell18} to study the impact of baryons on dark matter fractions, and in \textcolor{blue}{Terrazas et al. (in prep)} to probe the origin of quenched galaxies in the TNG model.

Each of the $\sim\,$100 TNG model variants simulates the exact same $25 {\rm Mpc}/h$ volume at a resolution approximately equivalent to the flagship TNG100-1. Individual halos can also therefore be cross-matched between the simulations, although the  statistics is necessarily limited by the relatively small volume. We plan to publicly release these variations in the near future, and encourage those interested to get in touch in the meantime.

Finally, we anticipate that ongoing and future simulation projects will also be released through this platform in the future. This includes upcoming IllustrisTNG `spinoff' projects, extending the scientific scope and goals beyond the original three boxes, as well as other simulations from other groups as well.

\subsubsection{API Functionality Expansion}

There is significant room for the development of additional features in the web-based API. In particular, for (i) on-demand visualization tasks, (ii) on-demand analysis tasks, and (iii) client-side, browser based tools for data exploration and visualization. We have two specific services which are anticipated to be developed in the near-term future and made available.

First, the on-demand generation of `zoom' initial conditions (ICs), for individual galaxies/halos, based on any object of interest selected from any simulation box. This will allow a user to select a sample of galaxies, perhaps in analogy to an observed sample, or with a peculiar type of assembly history, and obtain ICs for resimulation. Such resimulations could use other codes or galaxy formation models, or explore model parameter variations, to assess how such changes affected a particular galaxy/halo, or class of galaxies/halos. As IC generation will take several minutes at least, it does not fit within our current framework of `responses within a few seconds', and therefore requires a task-based work queue with delayed execution and subsequent notification (e.g. via email) of completion and the availability of new data products for download.
  
Second, the on-demand execution of longer running analysis tasks, with similar notification upon completion. Specifically, the ability to request SKIRT radiative-transfer calculations for specific galaxies/halos of interest, leveraging the development efforts of \cite{rodriguezgomez18}. Other expensive mocks, such as spectral HI \citep[with \textsc{MARTINI};][]{oman19} or x-ray datacubes, or intergalactic quasar absorption sightlines, can similarly be generated.

We welcome community input and/or contributions in any of these directions, or comments related to any aspect of the public data release of TNG.


\section{Architectural and Design Details} \label{sImplementation}

In the development of the original Illustris public data release, many design decisions were made, including technical details related to the release effort, based on expected use cases and methods of data analysis. \cite{nelson15b} discusses the architectural goals and considerations that we followed and continue to follow with the IllustrisTNG data release, and contrasts against other methodologies, as implemented in other astrophysics simulation data releases. We refer the reader to that paper and present only a few updates here.

\subsection{Usage of the Illustris Public Data Release}

Since its release, the original Illustris public data release has seen widespread adoption and use. To date, in the three and a half years since launch, 2122 new users have registered and made a total of 269 million API requests, including 2.7 million `mock FITS' file downloads. For the flagship Illustris-1 run, a total of 1390 full snapshots, 6650 group catalogs, and 180 merger trees have been downloaded. 26 million subhalo `cutouts' of particle-level data, and 3.1 million {\sc SubLink} merger tree extractions have been requested. The total data transfer for this simulation to date is $\simeq\,$2.15 PB. Roughly 3100 subbox snapshots of Illustris-1 have been downloaded. The next most accessed simulation is Illustris-3, likely because it is included in the getting started tutorials as an easy, lightweight alternative to Illustris-1. Since launch, there has been a nearly constant number of $\sim 100 - 120$ active users, based on activity within the last 30 days. 

To date, 163 publications have directly resulted from, or included analysis results from, the Illustris simulation. While early papers were written largely by the collaboration itself, recent papers typically do not involve members of the Illustris team, representing widespread public use of the data release. Of the 10 most recent papers published on Illustris, only one was from the team. Given the significantly expanded scope of TNG with respect to Illustris, as well as the relatively more robust and reliable physical model and outcomes, we expect that uptake and usage will be similarly broad.

\subsection{New JupyterLab Interface}

In the original Illustris data release, we promoted two ways to work with the data: either downloading large simulation data files directly (referred to above as `local data, local compute'), or by searching and downloading data subsets using functionality in the web-based API (`remote data, local compute'). Previously, the backend was focused solely on storage and data delivery, and did not have any system in place to allow guest access to compute resources which were local to the datasets themselves. For the TNG data release we have developed this functionality.

We label this newly introduced, third method of working with the data `remote data, remote computation'. Technically, we make use of JupyterHub to manage the instantiation of per-user JupyterLab instances. These are spawned inside containerized Docker instances \citep{merkel14} to isolate the user from the host systems -- Singularity \citep{kurtzer17} could be used in the future. Read-only mounts to the parallel filesystems hosting simulation data are provided, while the user home directory within the container is made persistent by volume mapping it onto the host. Resource limits on CPU, memory, and storage are controllable and will be adjusted during the initial phase of this service as needed.

The JupyterLab instances themselves provide a familiar environment for the development and execution of user analysis programs. Over the past few years there has been significant recent development on remote, multi-user, rich interfaces to computational kernels, and JupyterLab \citep[the successor of Jupyter, previously called IPython;][]{perez07} is a mature, full-featured solution we deploy. These instances are launched, on demand, inside the sand-boxed containers, through a web-based portal with authentication integrated into the existing user registration system of the data release. We anticipate that this will be a particularly interesting development for researchers who would otherwise not have the computational resources to use the simulation data for their science.

\subsection{Retiring the Relational Database}

In the original Illustris data release we noted that the read-only, highly structured nature of simulation output motivates different and more efficient approaches for data search, aggregation, processing, and retrieval tasks. The web-based API uses a representational state transfer architecture \citep[REST,][]{fielding00}, and in TNG we continue to employ a relational database on the backend, although we made a design decision never to expose such a database to direct user query.  

Looking forward, instead of bringing the object or group catalog data into a traditional database, one could employ a scheme such as bitmap indexing over HDF5, e.g. \href{http://www-vis.lbl.gov/Events/SC05/HDF5FastQuery/}{FastQuery} \citep{chou11,byna12}, possibly combined with a SQL-compatible query layer \citep{wang13}. In this case, the database would be used only to handle light meta-data -- fast index-accelerated search queries would be made directly against structured binary data on disk. This improvement would be largely transparent from the user perspective. Most obviously, it would remove a layer of complexity and the need to ingest of order billions of rows of group catalog data into a database. It would also enable a tighter coupling of search capabilities and on-disk data contents.

More efficient API standards such as \href{https://graphql.org/}{GraphQL} represent modern alternatives to REST, whereby users make specific, detailed requests to a single endpoint based on a well-defined query language and typed schema, rather than a number of generic requests to a diversity of endpoints. Resolving these declarative queries efficiently and directly on the simulation output data would unify many of these goals -- a clear target for future development.


\section{Summary and Conclusions} \label{sConclusions}

We have made publicly available data from the IllustrisTNG simulation project at the permanent URL:

\vspace{0.5em}
\begin{itemize}
\item[] \url{http://www.tng-project.org/data/}
\end{itemize}
\vspace{0.5em}

IllustrisTNG is a series of large-scale, cosmological simulations ideal for studying the formation and evolution of galaxies. 
The simulation suite consists of three volumes: TNG50, TNG100, and TNG300. Each flagship run is accompanied by lower-resolution realizations, and a dark-matter only analog of every simulation is also available.
The current data release includes TNG50, TNG100 and TNG300 in their entirety.
Full snapshots, group catalogs (both friends of friends halos and {\sc SubFind} subhalos), merger trees, high time-resolution subboxes, and many supplementary data catalogs are made available.
The highest resolution TNG300-1 includes more than ten million gravitationally bound structures, 
and the TNG100-1 volume contains $\sim$20,000 well-resolved galaxies at $z=0$ with stellar mass exceeding $10^{9} {\rm M}_\odot$.
The galaxies sampled in these volumes encompass a broad range of mass, type, environment and assembly history, and realize fully representative synthetic universes within the context of $\Lambda$CDM.

The total data volume produced by the Illustris[TNG] project is sizeable, $\sim$1.1 PB in total, all directly accessible online. 
We have developed several tools to make these data accessible to the broader community, without requiring extensive local computational resources.
In addition to direct data download, example scripts, web-based API access methods, and extensive documentation previously developed for the original Illustris simulation, we extend the data access functionality in several ways. 
Namely, with new on-demand visualization and analysis functionality, and with the remote JupyterLab-based analysis interface.
By making the TNG data publicly available, we aim to maximize the scientific return from the considerable computational resources invested in the TNG simulation suite.


\begin{backmatter}

\section*{Abbreviations}

Not applicable.

\section*{Declarations}

\subsection*{Availability of Data and Material}

This paper describes an explicit public data release of the relevant material.

\vspace{-1em}
\subsection*{Competing Interests}

The authors declare that they have no competing interests.

\vspace{-1em}
\subsection*{Funding}

Not applicable.

\vspace{-1em}
\subsection*{Author's Contributions}

DN has designed and carried out the data release, authored the presentation and data distribution websites, developed and maintains the technical infrastructure, and composed this document. VS, AP, RP and DN have carried out the TNG simulations. VS, LH, RW and AP have developed the TNG model with further input from SG, PT, MV, FM, RP and DN. VRG, LK, ML, and BD have contributed additional data products.

\vspace{-1em}
\subsection*{Acknowledgements} 

We thank Prof. Volker Springel, together with the Max Planck Computing and Data Facility (MPCDF) and the Max Planck Institute for Astrophysics (MPA) in Garching, Germany, for significant computational resources and ongoing support, without which the TNG public data release would not be possible. We also thank Prof. Lars Hernquist, together with the Research Computing group of Harvard University, for significant computational resources and continuous support, without which the original Illustris public data release and ongoing activities would not be possible.

In addition, we thank the Center for Computational Astrophysics and the Simons Foundation for a future commitment to host a full mirror of the Illustris[TNG] data, particularly Dylan Simon and the Scientific Computing Core at the Flatiron Institute for orchestration efforts. In addition, we thank Greg Snyder and the Mikulski Archive for Space Telescopes (MAST) at STScI for future plans to host relevant datasets based on TNG.

The authors would like to thank many additional people for contributing to analysis and understanding of the TNG simulations and their results: Maria Celeste Artale, Kelly Blumenthal, David Barnes, Martina Donnari, Elena D'Onghia, Min Du, Catalina Gomez, Hong Guo, Anshu Gupta, Melanie Habouzit, Chris Hayward, Ghandali Joshi, Guinevere Kauffmann, Davide Martizzi, Michelle Ntampaka, Ana-Roxana Pop, Gerg{\"o} Popping, Malin Renneby, Greg Snyder, Adam Stevens, Sandro Tacchella, Bryan Terrazas, Nhut Truong, Hannah {\"U}bler, Francisco Villaescusa-Navarro, Sarah Wellons, Po-Feng Wu, Dandan Xu, Kiyun Yun, Qirong Zhu, Elad Zinger, and Jolanta Zjupa.

In the execution of the primary simulations presented herein, the authors gratefully acknowledge the Gauss Centre for Supercomputing (GCS) for providing computing time for the GCS Large-Scale Projects GCS-ILLU (2014) and GCS-DWAR (2016) on the GCS share of the supercomputer Hazel Hen at the High Performance Computing Center Stuttgart (HLRS). GCS is the alliance of the three national supercomputing centres HLRS (Universit{\"a}t Stuttgart), JSC (Forschungszentrum J{\"u}lich), and LRZ (Bayerische Akademie der Wissenschaften), funded by the German Federal Ministry of Education and Research (BMBF) and the German State Ministries for Research of Baden-W{\"u}rttemberg (MWK), Bayern (StMWFK) and Nordrhein-Westfalen (MIWF). Additional simulations were carried out on the Hydra and Draco supercomputers at the Max Planck Computing and Data Facility (MPCDF, formerly known as RZG), as well as on the Stampede supercomputer at the Texas Advanced Computing Center through the XSEDE project AST140063. Some additional computations in this paper were run on the Odyssey cluster supported by the FAS Division of Science, Research Computing Group at Harvard University.


\bibliographystyle{bmc-mathphys} 
\bibliography{refs}
\nocite{label}

\end{backmatter}


\clearpage
\appendix
\onecolumn

\section*{Appendix A: Simulation Data Details}
\setcounter{table}{0}
\renewcommand*{\theHtable}{A.\arabic{table}}
\gdef\thetable{A.\arabic{table}}

\begin{table*}[htb!]
\footnotesize
  \caption{Details on the file organization for all twenty TNG runs, both baryonic and dark-matter only. We include the number of file chunks, the average size of a full snapshot and the corresponding group catalog, and an estimate of the total data volume of the simulation.}
  \label{table_chunks}
  \begin{center}
  \renewcommand{\arraystretch}{1.5}
    \begin{tabular}{llrcccc}
    \hline
   Run              & Alternate Name     & Total $N_{\rm DM}$ & N$_{\rm chunks}$ & Full Snapshot Size & Avg Groupcat Size & Total Data Volume \\ \hline\hline
   L35n270TNG       & TNG50-4            & 19,683,000         & 11   & 5.2 GB       & 20 MB  & 0.6 TB \\
   L35n270TNG\_DM   & TNG50-4-Dark       & 19,683,000         & 4    & 1.2 GB       & 10 MB  & 0.1 TB \\
   L35n540TNG       & TNG50-3            & 157,464,000        & 11   & 44 GB        & 130 MB & 7.5 TB \\
   L35n540TNG\_DM   & TNG50-3-Dark       & 157,464,000        & 4    & 9.4 GB       & 50 MB  & 0.6 TB \\
   L35n1080TNG      & TNG50-2            & 1,259,712,000      & 128  & 350 GB       & 860 MB & 18 TB  \\
   L35n1080TNG\_DM  & TNG50-2-Dark       & 1,259,712,000      & 85   & 76 GB        & 350 MB & 4.5 TB \\
   L35n2160TNG      & TNG50-1            & 10,077,696,000     & 680  & 2.7TB        & 7.2 GB & $\sim$320 TB \\
   L35n2160TNG\_DM  & TNG50-1-Dark       & 10,077,696,000     & 128  & 600 GB       & 2.3 GB & 36 TB  \\
   L75n455TNG       & TNG100-3           & 94,196,375         & 8    & 27 GB        & 110 MB & 1.5 TB \\
   L75n455TNG\_DM   & TNG100-3-Dark      & 94,196,375         & 4    & 5.7 GB       & 40 MB  & 0.4 TB \\
   L75n910TNG       & TNG100-2           & 753,571,000        & 56   & 215 GB       & 650 MB & 14 TB  \\
   L75n910TNG\_DM   & TNG100-2-Dark      & 753,571,000        & 8    & 45 GB        & 260 MB & 2.8 TB \\
   L75n1820TNG      & TNG100-1           & 6,028,568,000      & 448  & 1.7 TB       & 4.3 GB & 128 TB \\
   L75n1820TNG\_DM  & TNG100-1-Dark      & 6,028,568,000      & 64   & 360 GB       & 1.7 GB & 22 TB  \\
   L205n625TNG      & TNG300-3           & 244,140,625        & 16   & 63 GB        & 340 MB & 4 TB   \\
   L205n625TNG\_DM  & TNG300-3-Dark      & 244,140,625        & 4    & 15 GB        & 130 MB & 1 TB   \\
   L205n1250TNG     & TNG300-2           & 1,953,125,000      & 100  & 512 GB       & 2.2 GB & 31 TB  \\
   L205n1250TNG\_DM & TNG300-2-Dark      & 1,953,125,000      & 25   & 117 GB       & 810 MB & 7.2 TB \\
   L205n2500TNG     & TNG300-1           & 15,625,000,000     & 600  & 4.1 TB       & 14 GB  & 235 TB \\
   L205n2500TNG\_DM & TNG300-1-Dark      & 15,625,000,000     & 75   & 930 GB       & 5.2 GB & 57 TB  \\ \hline
    \end{tabular}
  \end{center}
\end{table*}

\section*{Appendix B: Web-Based API Examples}

\setcounter{table}{0}
\renewcommand*{\theHtable}{B.\arabic{table}}
\gdef\thetable{B.\arabic{table}}

By way of explicit example, the following are absolute URLs for the web-based API which encompass some of its functionality. The type of the request, as well as the data expected in return, should be relatively clear:

\begin{itemize}[leftmargin=1em]
\footnotesize
\setlength\itemsep{0.25em}
\item \url{www.tng-project.org/api/Illustris-2/}
\item \url{www.tng-project.org/api/Illustris-2/snapshots/68/}
\item \url{www.tng-project.org/api/Illustris-1/snapshots/135/subhalos/73664/}
\item \url{www.tng-project.org/api/Illustris-1/snapshots/80/halos/523312/cutout.hdf5?dm=Coordinates&gas=all}
\item \url{www.tng-project.org/api/Illustris-3/snapshots/135/subhalos?mass__gt=10.0&mass__lt=20.0}
\item \url{www.tng-project.org/api/Illustris-2/snapshots/68/subhalos/50000/sublink/full.hdf5}
\item \url{www.tng-project.org/api/Illustris-2/snapshots/68/subhalos/50000/sublink/mpb.json}
\item \url{www.tng-project.org/api/TNG100-2/snapshots/99/subhalos/50000/sublink/mpb.json}
\item \url{www.tng-project.org/api/TNG300-1/snapshots/99/subhalos?mass__gt=10.0&mass__lt=11.0}
\item \url{www.tng-project.org/api/Illustris-2/files/ics.hdf5}
\item \url{www.tng-project.org/api/Illustris-1/files/groupcat-135.5.hdf5}
\item \url{www.tng-project.org/api/Illustris-2/files/snapshot-135.10.hdf5?dm=all}
\item \url{www.tng-project.org/api/TNG100-1/snapshots/50/subhalos/plot.png?xQuant=mstar2_log&yQuant=ssfr}
\item \url{www.tng-project.org/api/TNG300-1/snapshots/7/subhalos/plot.png?xQuant=mstar2&yQuant=delta_sfms&cQuant=Z_gas}
\item \url{www.tng-project.org/api/TNG100-1/snapshots/99/halos/320/vis.png}
\item \url{www.tng-project.org/api/TNG100-2/snapshots/67/halos/0/vis.png?partType=gas&partField=temp}
\end{itemize}

A `getting started' guide for the web-based API is available in the online documention, and this includes a cookbook of common analysis tasks (available in Python, IDL, and Matlab). To give a sense of this method of analyzing TNG data, we include here four short examples, in Python. Each uses of the {\sc get()} helper function which performs the actual HTTP request, automatically parsing JSON-type returns.

\vspace{5mm}
\noindent\textbf{Task 1:} For TNG300-1 at $z=0$, get all the information available for the ID = 0 subhalo, print both its total mass and stellar half mass radius.
\begin{Verbatim}[commandchars=\\\{\}]
\PY{o}{\PYZgt{}\PYZgt{}}\PY{o}{\PYZgt{}} \PY{n}{url} \PY{o}{=} \PY{l+s}{\PYZdq{}}\PY{l+s}{http://www.tng\PYZhy{}project.org/api/TNG300\PYZhy{}1/snapshots/99/subhalos/0/}\PY{l+s}{\PYZdq{}}
\PY{o}{\PYZgt{}\PYZgt{}}\PY{o}{\PYZgt{}} \PY{n}{r} \PY{o}{=} \PY{n}{get}\PY{p}{(}\PY{n}{url}\PY{p}{)}
\PY{o}{\PYZgt{}\PYZgt{}}\PY{o}{\PYZgt{}} \PY{n}{r}\PY{p}{[}\PY{l+s}{\PYZsq{}}\PY{l+s}{mass}\PY{l+s}{\PYZsq{}}\PY{p}{]}
\PY{l+m+mf}{128335.0}

\PY{o}{\PYZgt{}\PYZgt{}}\PY{o}{\PYZgt{}} \PY{n}{r}\PY{p}{[}\PY{l+s}{\PYZsq{}}\PY{l+s}{halfmassrad\PYZus{}stars}\PY{l+s}{\PYZsq{}}\PY{p}{]}
\PY{l+m+mf}{130.065}
\end{Verbatim}

\vspace{3mm}
\noindent\textbf{Task 2:} For TNG100-1 at $z=2$, search for all subhalos with total mass $10^{12.1} {\rm M}_\odot < M < 10^{12.2} {\rm M}_\odot$ and print the {\sc Subfind} IDs of the first five results.
\begin{Verbatim}[commandchars=\\\{\}]
\PY{o}{\PYZgt{}\PYZgt{}}\PY{o}{\PYZgt{}} \PY{c}{\PYZsh{} convert from log solar masses to group catalog units}
\PY{o}{\PYZgt{}\PYZgt{}}\PY{o}{\PYZgt{}} \PY{n}{mass\PYZus{}min} \PY{o}{=} \PY{l+m+mi}{10}\PY{o}{*}\PY{o}{*}\PY{l+m+mf}{12,1} \PY{o}{/} \PY{l+m+mf}{1e10} \PY{o}{*} \PY{l+m+mf}{0.6774} 
\PY{o}{\PYZgt{}\PYZgt{}}\PY{o}{\PYZgt{}} \PY{n}{mass\PYZus{}max} \PY{o}{=} \PY{l+m+mi}{10}\PY{o}{*}\PY{o}{*}\PY{l+m+mf}{12.2} \PY{o}{/} \PY{l+m+mf}{1e10} \PY{o}{*} \PY{l+m+mf}{0.6774}
\PY{o}{\PYZgt{}\PYZgt{}}\PY{o}{\PYZgt{}} 
\PY{o}{\PYZgt{}\PYZgt{}}\PY{o}{\PYZgt{}} \PY{n}{params} \PY{o}{=} \PY{p}{\PYZob{}}\PY{l+s}{\PYZsq{}}\PY{l+s}{mass\PYZus{}\PYZus{}gt}\PY{l+s}{\PYZsq{}}\PY{p}{:}\PY{n}{mass\PYZus{}min}\PY{p}{,} \PY{l+s}{\PYZsq{}}\PY{l+s}{mass\PYZus{}\PYZus{}lt}\PY{l+s}{\PYZsq{}}\PY{p}{:}\PY{n}{mass\PYZus{}max}\PY{p}{\PYZcb{}}
\PY{o}{\PYZgt{}\PYZgt{}}\PY{o}{\PYZgt{}} 
\PY{o}{\PYZgt{}\PYZgt{}}\PY{o}{\PYZgt{}} \PY{c}{\PYZsh{} make the request}
\PY{o}{\PYZgt{}\PYZgt{}}\PY{o}{\PYZgt{}} \PY{n}{url} \PY{o}{=} \PY{l+s}{\PYZdq{}}\PY{l+s}{http://www.tng\PYZhy{}project.org/api/TNG100\PYZhy{}1/snapshots/z=2/subhalos/}\PY{l+s}{\PYZdq{}}
\PY{o}{\PYZgt{}\PYZgt{}}\PY{o}{\PYZgt{}} \PY{n}{subhalos} \PY{o}{=} \PY{n}{get}\PY{p}{(}\PY{n}{url}\PY{p}{,} \PY{n}{params}\PY{p}{)}
\PY{o}{\PYZgt{}\PYZgt{}}\PY{o}{\PYZgt{}} 
\PY{o}{\PYZgt{}\PYZgt{}}\PY{o}{\PYZgt{}} \PY{n}{ids} \PY{o}{=} \PY{p}{[} \PY{n}{subhalos}\PY{p}{[}\PY{l+s}{\PYZsq{}}\PY{l+s}{results}\PY{l+s}{\PYZsq{}}\PY{p}{]}\PY{p}{[}\PY{n}{i}\PY{p}{]}\PY{p}{[}\PY{l+s}{\PYZsq{}}\PY{l+s}{id}\PY{l+s}{\PYZsq{}}\PY{p}{]} \PY{k}{for} \PY{n}{i} \PY{o+ow}{in} \PY{n+nb}{range}\PY{p}{(}\PY{l+m+mi}{5}\PY{p}{)} \PY{p}{]}
\PY{o}{\PYZgt{}\PYZgt{}}\PY{o}{\PYZgt{}} \PY{n}{ids}
\PY{p}{[}\PY{l+m+mi}{13845}\PY{p}{,} \PY{l+m+mi}{16799}\PY{p}{,} \PY{l+m+mi}{23224}\PY{p}{,} \PY{l+m+mi}{24400}\PY{p}{,} \PY{l+m+mi}{12718}\PY{p}{]}
\end{Verbatim}

\vspace{5mm}
\noindent\textbf{Task 11:} Download the entire TNG300-1 $z=0$ snapshot including \textit{only the positions, masses, and metallicities of stars} (in the form of 600 HDF5 files). 
In this example, since we only need these three fields for stars only, we can reduce the download and storage size from $\sim$4.1\,TB to $\sim$20\,GB.
\begin{Verbatim}[commandchars=\\\{\}]
\PY{o}{\PYZgt{}\PYZgt{}}\PY{o}{\PYZgt{}} \PY{n}{base\PYZus{}url} \PY{o}{=} \PY{l+s}{\PYZdq{}}\PY{l+s}{http://www.tng\PYZhy{}project.org/api/TNG300\PYZhy{}1/}\PY{l+s}{\PYZdq{}}
\PY{o}{\PYZgt{}\PYZgt{}}\PY{o}{\PYZgt{}} \PY{n}{sim\PYZus{}metadata} \PY{o}{=} \PY{n}{get}\PY{p}{(}\PY{n}{base\PYZus{}url}\PY{p}{)}
\PY{o}{\PYZgt{}\PYZgt{}}\PY{o}{\PYZgt{}} \PY{n}{params} \PY{o}{=} \PY{p}{\PYZob{}}\PY{l+s}{\PYZsq{}}\PY{l+s}{stars}\PY{l+s}{\PYZsq{}}\PY{p}{:}\PY{l+s}{\PYZsq{}}\PY{l+s}{Coordinates,Masses,GFM\PYZus{}Metallicity}\PY{l+s}{\PYZsq{}}\PY{p}{\PYZcb{}}
\PY{o}{\PYZgt{}\PYZgt{}}\PY{o}{\PYZgt{}} 
\PY{o}{\PYZgt{}\PYZgt{}}\PY{o}{\PYZgt{}} \PY{k}{for} \PY{n}{i} \PY{o+ow}{in} \PY{n+nb}{range}\PY{p}{(}\PY{n}{sim\PYZus{}metadata}\PY{p}{[}\PY{l+s}{\PYZsq{}}\PY{l+s}{num\PYZus{}files\PYZus{}snapshot}\PY{l+s}{\PYZsq{}}\PY{p}{]}\PY{p}{)}\PY{p}{:}
\PY{o}{\PYZgt{}\PYZgt{}}\PY{o}{\PYZgt{}}     \PY{n}{file\PYZus{}url} \PY{o}{=} \PY{n}{base\PYZus{}url} \PY{o}{+} \PY{l+s}{\PYZdq{}}\PY{l+s}{files/snapshot\PYZhy{}99.}\PY{l+s}{\PYZdq{}} \PY{o}{+} \PY{n+nb}{str}\PY{p}{(}\PY{n}{i}\PY{p}{)} \PY{o}{+} \PY{l+s}{\PYZdq{}}\PY{l+s}{.hdf5}\PY{l+s}{\PYZdq{}}
\PY{o}{\PYZgt{}\PYZgt{}}\PY{o}{\PYZgt{}}     \PY{n}{saved\PYZus{}filename} \PY{o}{=} \PY{n}{get}\PY{p}{(}\PY{n}{file\PYZus{}url}\PY{p}{,} \PY{n}{params}\PY{p}{)}
\PY{o}{\PYZgt{}\PYZgt{}}\PY{o}{\PYZgt{}}     \PY{k}{print}\PY{n}{(saved\PYZus{}filename)}
\end{Verbatim}

\end{document}